\documentclass[superscriptaddress,unsortedaddress,
 amsmath,amssymb,
 aps,
]{revtex4-2}

\usepackage{graphicx}
\usepackage{dcolumn}
\usepackage{bm}
\usepackage{physics}
\usepackage{lipsum}
\usepackage{siunitx}
\usepackage{color}
\usepackage{subcaption}
\usepackage{tikz,pgfplots}
\usepgfplotslibrary{groupplots,dateplot}
\usetikzlibrary{patterns,shapes.arrows}
\usepackage{makecell}
\pgfplotsset{compat=1.17}
\usepackage{hyperref} 

\usetikzlibrary{patterns,shapes.arrows, arrows.meta,bending, shapes,calc,fadings,decorations.pathreplacing,positioning,arrows.meta}
\pgfplotsset{compat=newest}

\usepackage{sansmath}
\tikzset{>=stealth,
OptimumStyle/.style={align=center,anchor=east,rotate=90,font=\scriptsize}
}
\pgfplotsset{
axis lines = left,
every axis plot/.append style={line width=2pt},
}
\usepackage{forest}
\forestset{
    .style={
        for tree={
            base=bottom,
            child anchor=north,
            align=center,
            s sep+=1cm,
    straight edge/.style={
        edge path={\noexpand\path[\forestoption{edge},thick,-{Latex}]
        (!u.parent anchor) -- (.child anchor);}
    },
    if n children={0}
        {tier=word, draw, thick, rectangle}
        {draw, diamond, thick, aspect=2},
    if n=1{%
        edge path={\noexpand\path[\forestoption{edge},thick,-{Latex}]
        (!u.parent anchor) -| (.child anchor) node[pos=.2, above] {Y};}
        }{
        edge path={\noexpand\path[\forestoption{edge},thick,-{Latex}]
        (!u.parent anchor) -| (.child anchor) node[pos=.2, above] {N};}
        }
        }
    }
}
\tikzset{
  green arrow/.style={
    midway,green,sloped,fill, minimum height=2cm, single arrow, single arrow head extend=.5cm, single arrow head indent=.25cm,xscale=0.3,yscale=0.15,
    allow upside down
  },
  yellow arrow/.style={
    midway,yellow,sloped,fill, minimum height=2cm, single arrow, single arrow head extend=.5cm, single arrow head indent=.25cm,xscale=0.3,yscale=0.15,
    allow upside down
  },
  black arrow/.style 2 args={-stealth, shorten >=#1, shorten <=#2},
  black arrow/.default={1mm}{1mm},
  tree box/.style={draw, rounded corners, inner sep=1em},
  node box/.style={white, draw=white, text=black, rectangle, rounded corners},
}

\begin{document}

\title{Predicting Solid State Material Platforms for Quantum Technologies}

\author{Oliver Lerst{\o}l Hebnes}
\affiliation{Sopra Steria, Information Technology and Services, N-4020 Stavanger, Norway}
\affiliation{Department of Physics and Center for Computing in Science Education, University of Oslo, N-0316 Oslo, Norway}

\author{Marianne Etzelm\"uller Bathen}
\email{bathen@aps.ee.ethz.ch}
\affiliation{Advanced Power Semiconductor Laboratory, ETH Z{\"u}rich, 8092  Z{\"u}rich,  Switzerland}

\author{{\O}yvind Sigmundson Sch{\o}yen}
\affiliation{Department of Physics and Center for Computing in Science Education, University of Oslo, N-0316 Oslo, Norway}

\author{Sebastian G. Winther-Larsen}
\affiliation{Menon Economics, N-0369 Oslo, Norway}
\affiliation{Department of Physics and Center for Computing in Science Education, University of Oslo, N-0316 Oslo, Norway}

\author{Lasse Vines}
\affiliation{Department of Physics and Center for Materials Science and Nanotechnology, University of Oslo, N-0316 Oslo, Norway}

\author{Morten Hjorth-Jensen}
\affiliation{Department of Physics and Astronomy and Facility for Rare Ion Beams, Michigan State University, East Lansing, MI 48824, USA}
\affiliation{Department of Physics and Center for Computing in Science Education, University of Oslo, N-0316 Oslo, Norway}

\begin{abstract}

Semiconductor materials provide a compelling platform for quantum technologies (QT), and the properties of a vast amount of materials can be found in databases containing information from both experimental and theoretical explorations. However, searching these databases to find promising candidate materials for quantum technology applications 
is a major challenge. Therefore, we have developed a framework for the automated discovery of semiconductor host platforms for QT using material informatics and machine learning methods, resulting in a dataset consisting of over $25.000$ materials and nearly $5000$ physics-informed features. Three approaches were devised, named the Ferrenti, extended Ferrenti and the empirical approach,  to label data for the supervised machine learning (ML) methods logistic regression, decision trees, random forests and gradient boosting. 
We find that of the three, the empirical approach relying exclusively on findings from the literature predicted substantially fewer candidates than the other two approaches with a clear distinction between suitable and unsuitable candidates when comparing the two largest eigenvalues in the covariance matrix. 
In contrast to expectations from the literature and that found for the Ferrenti and extended Ferrenti approaches focusing on band gap and ionic character, the ML methods from the empirical approach highlighted features related to symmetry and crystal structure, including bond length, orientation and radial distribution, as influential when predicting a material as suitable for QT. 
All three approaches and all four ML methods agreed on a subset of $47$ eligible candidates 
of $8$ elemental, $29$ binary, and $10$ tertiary compounds, and provide a basis for further material explorations towards quantum technology.

\end{abstract}

\pacs{02.70.Ss, 31.15.A-, 31.15.bw, 71.15.-m, 73.21.La}

\maketitle

\section*{Introduction}
Quantum technologies (QT) based on solid state platforms have attracted a lot of attention during the last few years. 
Among the promising applications that are already available we find important breakthroughs such as in vivo sensing of magnetic fields in cells \cite{Lesage_2013}, secure communication over large distances by separation of entangled photons \cite{Ursin2007} and, notably, various quantum information processing prototypes and architectures \cite{Arute_2019}.  
Quantum computers are in high demand to meet the increasing need for computing power to solve complex and high-dimensional scientific problems. 
Recent advances have highlighted the potential of quantum information processors to outperform state-of-the art high-performance computing facilities. Indeed, Google's $53$ qubit quantum computer based on superconducting electronics solved a computational problem that was beyond the capabilities of a $200.000$ core supercomputer \cite{Arute_2019}. Most recently, IBM announced its $127$ qubit quantum processor~\cite{IBM2021}. Simultaneously, the concepts of entanglement and teleportation may eventually facilitate advanced quantum communication protocols such as quantum cryptography and the quantum internet \cite{quantum-internet-kimble}, spurring further investigations into technologies based on quantum mechanics.

\begin{figure}[t]
    \centering
    \includegraphics[width=0.9\textwidth]{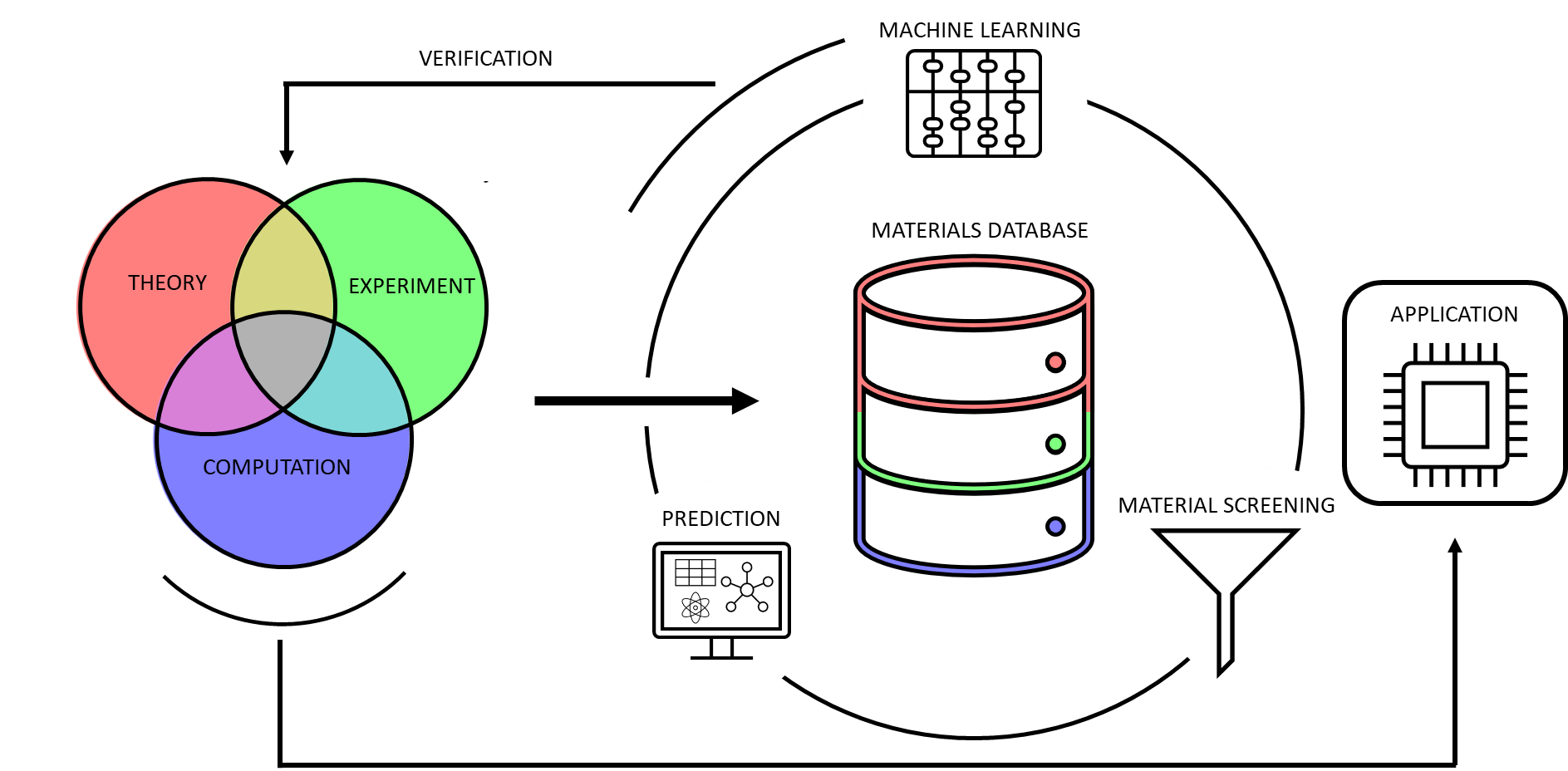}
    \caption{Schematic of an example workflow in material informatics. Results from theory, experiment and computation are fed into material databases (arrow pointing to the center). A cycle involving material screening, machine learning and predictions leads to knowledge gain and ultimately applications in fields such as clean energy and quantum technology. 
    }
    \label{fig:ht-workflow}
\end{figure}

Several platforms are available for the development of quantum technologies, but the materials and fabrication technologies are less mature than those for, e.g., classical computers and sensors. 
An important concern in this context is that of scalability. 
For example, the best performing quantum computer prototypes available today rely on superconducting electronics that require millikelvin temperatures to operate, with the stability of interactions between qubits being an important issue. Instead, semiconductors are emerging as a promising alternative platform, offering competitive characteristics combined with the possibility of room temperature operation and mature and scalable material processing and fabrication.  

Quantum technologies based on semiconductors rely on either defects or quantum dots where the latter can be of the self-assembled or nanostructured type \cite{Aharonovich_2016}. 
Semiconductor defects can act as single-photon emitters or spin centers and are compatible with the three main QT categories of computing, communication and sensing \cite{Awschalom_2018}. 
These characteristics are most often found for the case of defects that introduce deep energy levels into the semiconductor band gap \cite{Weber2010}. So-called deep level defects can trap charge carriers in localized states that are essentially isolated from the surroundings, making them highly suitable for QT due to, e.g., 
indistinguishable  
single-photon emission and long spin coherence times. 
The most well-known quantum compatible defect is the negatively charged nitrogen-vacancy (NV) center in diamond \cite{Doherty_2013}, but silicon carbide (SiC) and the various quantum emitters therein are strong contenders for quantum communication purposes especially due to the favorable emission wavelength region in the near infrared coupled with more mature material processing and fabrication (see, e.g., Refs.~\cite{Castelletto_2015,Son2020,Bathen2021}). 
However, semiconductor based QT is still in the early stages, and the issues left to address include identification of suitable host materials and candidate defects, and scalable and reproducible quantum device fabrication. 
Furthermore, we lack a complete understanding of the requirements for a semiconductor material to manifest quantum compatible properties,  
and the selection of known quantum compatible host materials is slim \cite{Atatuere2018,Zhang2020}. 

The majority of discoveries of QT compatible characteristics in semiconductors has so far happened by serendipity, and there is an urgent need for a better and more systematic understanding of which material requirements must be met for QT compatible characteristics like single-photon emission and single spin control to manifest. In this context, a framework for dedicated materials search and analysis is needed. 

The fourth science paradigm of big data driven science reveals the potential of targeted search for promising material systems in which to expect QT compatible properties. 
Rather than searching through a host of signals for those that match our criteria, we aim to \textit{predict} which materials and signatures we should target for more detailed studies, following the framework illustrated in Fig.~\ref{fig:ht-workflow}. 
This is made feasible by the availability of databases containing material properties for a wide range of different systems. In this work, the data in question are provided by bulk density functional theory (DFT) calculations to obtain the ground state properties of different elements and compounds. Combined with machine learning (ML) methods we provide a path towards precise classification of candidate materials. The inclusion of ML methods follows recent trends in applications of statistical learning, data science and machine learning for scientific discoveries, see for example Refs.~\cite{deiana2021,Carleo2019}. 

Here we provide a framework for the data mining and automated discovery of promising solid-state semiconductor hosts for quantum emitters and spin centers using targeted database search and ML methods combined with knowledge from the field. 
Analyzing the output of the ML methods reveals that, given a suitable initial set of labeled materials for training and testing, it is possible to discern the physical mechanisms that govern a material's suitability for quantum applications.  
In this framework we do not distinguish between the specific mechanism giving rise to properties such as single-photon emission and long spin coherence times (e.g., semiconductor defects or quantum dots); instead, we attempt to target all materials that may accommodate the desired characteristics.  
The methodology developed herein can be modified for other material types and application areas provided that high quality databases containing relevant theoretical and/or experimental data are available. 

The developed procedure relies on data extraction from the databases Materials Project \cite{Jain2013,Jain2018}, the Open Quantum Materials Database (OQMD) \cite{Saal2013, Kirklin2015}, JARVIS-DFT \cite{Choudhary2020}, AFLOW \cite{Curtarolo2012, Curtarolo2012a, Calderon2015} and AFLOW-ML \cite{Isayev2017}. 
The Matminer Python library for data mining \cite{Ward2018} was then used for material analysis to featurize the extracted data. An important aspect of our work is the database building and pertinent development of the datasets for the ML methods. We have developed three different approaches to data mining: (i) \emph{the Ferrenti approach} which is similar to that proposed by \citeauthor{Ferrenti2020} \cite{Ferrenti2020}, (ii) \emph{the extended Ferrenti approach} and (iii) \emph{the empirical approach}. We find that the two first methodologies base their data extraction protocols on broad material descriptors, leading to large sets of potentially suitable candidates \cite{Mehta2019,Hastie2009}. The empirical approach, on the other hand, relies on including materials with experimentally proven advantageous characteristics in the training set and therefore yields a narrower set of possible candidates. The three resulting sets of labeled data were then analyzed with the four supervised ML methods logistic regression, decision trees, random forests and gradient boosting \cite{Hastie2009,Murphy2012}, yielding $214$ predicted candidates that were common between all approaches and ML methods.  
Example materials include ZnGeP$_2$, CdS, BP, BC$_2$N, GeC and InP. Focused theoretical and experimental studies, using methods such as, e.g., density functional theory, photoluminescence spectroscopy and optically detected magnetic resonance, are needed to verify the predictions that quantum properties may manifest as a result of defects or nanostructures in the above listed materials. 
Importantly, our findings also reveal which material properties are weighted by the ML methods upon predicting a material as suitable for QT applications, thereby opening up for new  discoveries in the field of quantum technologies.

\section*{Results}

\subsection*{Information Flow} 

\begin{figure}[t]
    \centering
    \includegraphics[width=0.9\textwidth]{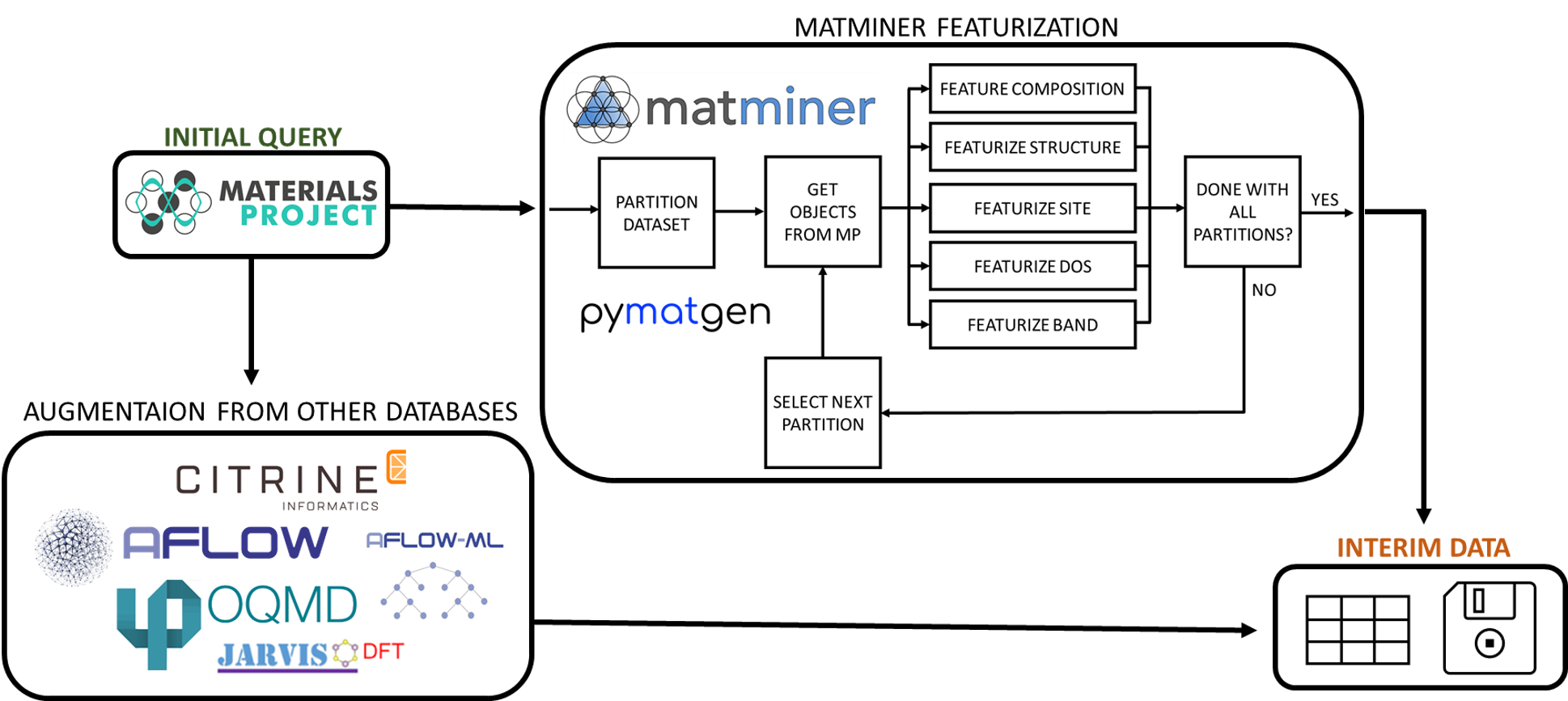}
    \caption{The project workflow starting from an initial Materials Project (MP) query, and ending with a featurized dataset with entries from several other databases. 
    To limit the memory and computational usage, the data are partitioned into smaller subsets where the respective Pymatgen objects (Pymatgen is a robust, open-source Python library for materials analysis \cite{pymatgen}) are obtained through a query to be used in the following featurization steps. This process is repeated iteratively until all the data has been featurized. \emph{DOS} refers to density of states and \emph{band} to the electronic band structure. 
    }
    \label{fig:flowchart}
\end{figure}

The information stream in this work can be regarded as many inter-connected modular parts. 
The initial step for gathering material data and building features is visualized by the outer flowchart in Fig.~\ref{fig:flowchart} (Jupyter notebooks containing the full
workflow can be found at \cite{Ohebbi2021}).
We start by extracting all entries in the Materials Project (MP) database  \cite{Jain2013,Jain2018} that match a specific query. 
The MP database contains ground state properties of different materials that are computed using density functional theory. The DFT calculations in the database were performed using the Vienna {\em ab initio} simulation package (VASP) \cite{Kresse1996} and the Perdew-Burke-Ernzerhof (PBE) \cite{Perdew1996} exchange-correlation functional to calculate the electronic structure of the materials. 
We note that despite being immensely successful in describing a number of material properties, PBE is widely known for underestimating the band gap of semiconductors \cite{Freysoldt2014}. Therefore, not all properties predicted using the PBE functional are reliable, and the band gaps in particular cannot be trusted in absolute terms. Nonetheless, the functional is in wide use due to the combination of reasonable accuracy and high computational throughput, and is usually considered to be reliable for large-scale trends in semiconductor properties. 

The conditions for the initial MP query are that the materials must derive from the Inorganic Crystal Structure Database (ICSD) and have a band gap wider than $0.1$~eV to exclude metallic compounds. The ICSD is the world's largest database for completely identified inorganic crystal structures \cite{Allen1987,Zagorac2019}. In a parallel step, entries that are deemed similar to the entries from the initial query are extracted from the databases OQMD  \cite{Saal2013,Kirklin2015}, JARVIS-DFT \cite{Choudhary2020}, AFLOW \cite{Curtarolo2012, Curtarolo2012a, Calderon2015}, AFLOW-ML \cite{Isayev2017} and the Citrination platform \cite{OMaraJordan2016MDIA}. The results of these steps are combined into a dataset for further analysis. 

After material extraction we apply tools from the open-source library Matminer \cite{Ward2018} to generate thousands of features from the data. We will refer to this process as featurization. A schematic visualization of the featurization process in Matminer is shown in  Fig.~\ref{fig:flowchart} 
and focuses on a material's composition,  structure, atomic sites, density of states and band structure. 
The $39$ featurizers (each generates several features) selected as material descriptors in this work are summarized and described in the Supplementary Information at \cite{supplementary}. The selection of features was kept rather wide as we do not want to make {\em ex ante} assumptions on which features best describe a solid state material platform for quantum technology. 

The constructed dataset encompasses compounds formed by a plethora of combinations of surfaces, interfaces, nanostructures, compositions and structures, but this complexity is not necessarily reflected in the material descriptors. 
Furthermore, we have utilized data obtained from high-throughput density functional theory calculations. Indeed, there are possible errors associated with every step, starting from an initial calculation, adding of data in the database, gathering of data, featurization of data, preprocessing of data, data mining, and finally training a model and making a prediction. Unfortunately, if an error has happened in the first part of the process, the error follows the entire process and will get increasingly harder to detect. Thus, our results depend on the quality of data in the databases we employ. 

\subsection*{Data Mining}
The complete dataset consists of $25.000$ materials. A subset of these materials is labeled into either suitable or unsuitable candidates for QT while the remainder will stay unlabeled. The labeled data are then grouped into training and test sets for the ML methods. 

Several challenges accompany the labeling process of labeling materials according to their suitability for QT. 
Although QT compatible properties are becoming increasingly well studied for the case of, e.g., nitrogen-vacancy centers in diamond, single-photon emitters in silicon carbide and quantum dot (QD) structures \cite{Doherty_2013,Bathen2021,Aharonovich_2016}, relatively few candidate materials are known to be suitable \cite{Atatuere2018,Zhang2020}. An additional consideration is that the physical mechanisms promoting favorable properties are not fully understood. 
Conversely, defining materials as unsuitable candidates for QT is in many ways equally difficult, as the mechanisms preventing quantum compatible characteristics to manifest are also unknown. 
The strategy for selection of unsuitable candidates is thus that the negation of the criteria we use to select suitable candidates should give unsuitable candidates.
A side-effect of this selection is that the criteria for unsuitable candidates can become equally, or more, restrictive compared to the criteria for suitable candidates. 
We will then find ourselves in a situation with skewed data sets were we have fewer unsuitable than suitable candidates. 
This method for selecting unsuitable candidates has shortcomings but will serve as a starting point and demonstration for how the labeling procedure could be improved. 
Below we follow three separate procedures for labeling materials as suitable or unsuitable candidates for QT. 

\subsubsection*{The Ferrenti approach}
The first approach to labeling a selection of materials in the dataset is based on the criteria proposed by \citeauthor{Ferrenti2020} \cite{Ferrenti2020}. 
They suggest a data mining process consisting of four stages by systematically evaluating the suitability of host materials taken from the Materials Project. In this framework, we label \textbf{suitable} candidates by the following steps: 
\begin{enumerate}
    \item Include materials that;
    \begin{itemize}
        \item contain elements with more than $50 \ \%$ natural abundance of spin zero isotopes,
        \item crystallize in non-polar space groups,
        \item are present in the ICSD database, 
        and
        \item are calculated to be nonmagnetic. 
    \end{itemize}
    \item Pragmatically remove toxic, radioactive and otherwise ``difficult'' materials;
    \begin{itemize}
        \item exclude Th, U, Cd and Hg because they are radioactive and/or toxic in the most stable forms,
        \item exclude any rare-earth metals (because of the difficulty of obtaining pure materials free of isotopes with nuclear spin) and noble gases (due to the lack of stable solid phases),
        \item exclude transition metal elements with unpaired electrons like Fe and Ni because of their paramagnetism; Ru and Os are also excluded because they only exist in the dataset as complex cluster structures. 
    \end{itemize}
    \item Include only materials with a band gap larger than $0.5$~eV calculated using DFT and the PBE functional. The value of $0.5$~eV was chosen to match that typically predicted for silicon by PBE-level DFT calculations. 
    \item Ensure that the energy above hull is less than $0.2$~eV per atom.
\end{enumerate}

These inclusion criteria are based on the work of \citeauthor{Weber2010} \cite{Weber2010} and targeted primarily at semiconductors that can host deep level defects with spin qubit capabilities. In this context, long spin coherence times are needed, and the possibility of having an environment free of nuclear spins and permanent magnetism is therefore desirable. 
Moreover, non-polar materials are assumed to be preferable to obtain defect based single-photon emitters with sharp zero-phonon lines and indistinguishable single-photon emission.  
Transition metal elements are eliminated if they have unpaired electrons because the presence of permanent electric dipole moments may have a detrimental impact on the optical coherence of defect emission. 
Finally, the energy above hull requirement is included to ensure that the selected compounds are thermodynamically stable. 

Next, \textbf{unsuitable} candidates are labeled according to the reverse requirements of the above; as materials in the ICSD database \cite{Allen1987,Zagorac2019} 
that crystallize in polar space groups, are calculated to be magnetic and have a band gap larger than $0.1$~eV in the MP database (to exclude metals but include lower-band gap semiconductors). 
The resulting set of labeled materials contains $1581$ materials where $35 \ \%$
are labeled as unsuitable and 
$65 \ \%$ as suitable for QT applications.
Here we see the first example of how the criteria for labeling unsuitable candidates are more restrictive than those for the suitable candidates.

\subsubsection*{The extended Ferrenti approach}
Here, we adjust the Ferrenti approach to expand the data labeling process beyond practical  considerations. The second approach is therefore named \emph{the extended Ferrenti approach} and involves removing stage two from the approach above. Moreover, we include a few additional elements that have shown promising properties but were initially excluded due to the lack of spin zero isotopes. 

The following steps constitute the process of labeling \textbf{suitable} candidates in the \emph{extended Ferrenti approach}:
\begin{enumerate}
    \item Include materials that; 
    \begin{itemize}
        \item contain elements where more than half have a natural abundance of spin zero isotopes, including Al, P, Ga, As, B and N, 
        \item crystallize in non-polar space groups,
        \item are present in the ICSD,
        \item are calculated to be nonmagnetic. 
    \end{itemize}
    \item Only keep materials that have a band gap larger than $1.5$~eV in the MP database. The higher band gap requirement (as compared to the Ferrenti approach) is included here to avoid labeling an unfeasibly large number of materials. 
    \item Ensure that the energy above hull is less than $0.2$~eV per atom. 
\end{enumerate}

For \textbf{unsuitable} candidates, we implement the same strategy as for the first Ferrenti approach. The result is a somewhat unbalanced set of labeled materials, with up to $75 \ \%$ of the materials found in the suitable group. However, the labeled data encompasses $78 \ \%$ more materials than  the Ferrenti approach, consisting of $2141$ suitable and $684$ unsuitable materials.

\subsubsection*{The empirical approach}
The third approach differs substantially from the others in terms of material labeling, as we apply knowledge from the field (see for instance Refs.~\cite{Atatuere2018,Toth2019,Zhang2020,Son2020} for an overview) to pick our promising material host candidates.  
In other words, we restrict ourselves to materials where quantum compatible properties have been either experimentally demonstrated or theoretically predicted. 
We therefore name this scheme \emph{the empirical approach}. Due to the concern of ending up with too small datasets for training and testing of the ML methods, we will also include some  materials that are promising in the sense that they have suitable properties for accommodating deep level defects that can potentially exhibit quantum effects --- even though such effects may not yet have been shown. These material choices were motivated by the criteria and predictions for QT compatible deep level defects defined by \citeauthor{Weber2010} \cite{Weber2010}. 
We have also included materials where properties like single-photon emission and spin manipulation have been observed but attributed to excitonic effects or quantum dots (formed by self-assembly or lithographic structuring) rather than being defect related.

Table~\ref{tab:qt-materials} contains an overview of known semiconductor materials with demonstrated quantum compatible characteristics. The table forms the basis for picking suitable candidates for the empirical approach. The properties being studied arise from mechanisms related to, e.g., point defects, bound excitons, and both self-assembled and lithographically structured quantum dots and nanostructures such as 2D materials. 
Quantum emission signatures have been assigned to specific defects in both diamond and SiC, but for most other materials, secure identification of the responsible defects or structure related mechanism is still lacking.  

The strategy for picking \textbf{suitable} candidates in the empirical approach can be summed up as;  
\begin{enumerate}
    \item Select candidate materials that match the formulas in  Table~\ref{tab:qt-materials}, or the formulas ZnSe, AlP, GaP, AlAs, ZnTe, CdS \cite{Weber2010} and SiGe \cite{Hardy2019}, as these materials have been predicted to behave as suitable quantum hosts based on favorable properties such as a wide band gap and low spin-orbit coupling.  
    \item Ensure that the candidates are present in the ICSD.  
    \item Perform a manual screening for appropriate crystallographic structures. 
\end{enumerate}

\begin{table}[b]
    \centering 
    \caption{Overview of materials and defects that have been demonstrated to exhibit quantum  compatible characteristics such as single-photon emission and coherent spin manipulation. The subscript denotes lattice site and $V$ refers to a vacancy. }
    \begin{tabular}{c|c|c|c}
    Material & Band gap (eV) & Defect candidates & References \\
    \hline
    Diamond  & $5.5$  & N$_\mathrm{C}V_\mathrm{C}$, Si$_\mathrm{C}V_\mathrm{C}$, Ge$_\mathrm{C}V_\mathrm{C}$ & \cite{Taylor2008,Balasubramanian_2009,Barclay2011,Gordon2013,Rogers_2014,Bhaskar_2018} \\ 
    SiC & $2.2$-$3.3$ & $V_\mathrm{Si}$, $V_\mathrm{Si}V_\mathrm{C}$, C$_\mathrm{Si}V_\mathrm{C}$, N$_\mathrm{C}V_\mathrm{Si}$ & \cite{Widmann2014,Christle_2015,Castelletto_2014,Zargaleh_2018}  \cite{Weber2010, Son2020, Falk2013} \\ 
    Si & $1.1$ & P, G, unidentified & \cite{Muhonen_2014,Durand_2020,Redjem2020} \\ 
    (2D) \textit{h}-BN & $6.0$ & Unidentified defects & \cite{Tran_2016,Tran_2016b,Hayee_2020} \\ 
    (2D) MoS$_2$, WSe$_2$, WS$_2$ & $<2.5$~eV & Bound excitons & \cite{Toth2019} \\
    ZnO & $3.4$ & Unidentified defects & \cite{Morfa2012} \\ 
    ZnS & $3.6$ (zincblende) & Unidentified defects & \cite{Stewart2019} \\ 
    GaAs & $1.4$ & Quantum dots & \cite{Bluhm2010} \\ 
    GaN & $3.4$ & Quantum dots, unidentified defects & \cite{Roux2017,Berhane2018} \\
    AlN & $6.0$ & Unidentified defects & \cite{Xue2020}\\
    \end{tabular}
    \label{tab:qt-materials}
\end{table} 

After the first stage of picking candidates we are left with a list of $202$ matching formulas which includes $12$ entries that have a band gap of less than $0.4$~eV. These $12$ entries are calculated to be thermodynamically unstable in terms of the energy above hull, and will decompose into other materials in the list --- incidentally, the resulting structures all have calculated band gaps that are substantially larger than $0.5$~eV. We include all of the $202$ structures apart from C (mp-$568410$) which is a metal according to AFLOW-ML.

Entries matching the formulas C, SiC, BN, MoS$_2$, WSe$_2$ and WS$_2$ (quantum compatible characteristics have been demonstrated) were manually screened to see whether they have a matching structure to the respective candidates discussed earlier and summarized in Table~\ref{tab:qt-materials}. 
For carbon, we admit three-dimensional diamond-like structures as explicitly stated in the column tags from the Materials Project. Additionally, we find several two-dimensional structures of carbon with a large band gap ($>1.5$~eV) among the data. We add these as suitable candidates. Complex structures (e.g., C$_{28}$, C$_{48}$ and C$_{60}$) were moved to the test set in our machine learning studies. For SiC we admitted all entries which included the $2$H, $3$C, $4$H, $6$H and $15$R polytypes. Concerning BN, MoS$_2$, WSe$_2$ and WS$_2$, we only admit two-dimensional structures. For non-matching structures not mentioned so far, we move them to the test set to see if they will be predicted as suitable or not by the ML methods that are applied in a later stage.

The materials AlP, GaP, AlAs, ZnTe and CdS were manually screened for tetrahedrally coordinated structures, and have been included since \citeauthor{Weber2010} \cite{Weber2010} identified them as potentially promising candidates due to suitable material properties. 
We note that only tetrahedrally coordinated structures of the given formulas were labeled as suitable after imposing a band gap restriction of $0.5$~eV. 

Following the three screening steps in the empirical approach, a total of $187$ entries were labeled as suitable candidates for the training set. 
Notably, the candidates labeled as suitable in the empirical approach contain only compounds that are either elementary (unary) or binary. 
Since the set of labeled data constituting suitable candidates in the empirical approach contains rather few entries, we add $400$ candidates that are labeled as \textbf{unsuitable}. These are picked at random from the pool of unsuitable candidates from the two previous approaches, in addition to those that were marked as unsuitable during the manual screening process. 

\subsubsection*{Comparing the approaches}
The three different approaches to data mining place a particular emphasis on each of their specific goals. The Ferrenti approach relies on choosing only elements with spin zero isotopes and compounds that crystallize in nonpolar spacegroups together with practical filters, while the extended Ferrenti approach allows a larger variety of elements and removes the practical reasons for excluding candidate materials. Thus, the first approach targets a more narrow prediction space than the second approach does. However, the most selective approach is the empirical one, relying exclusively on previous findings rather than assuming which material properties are influential.

Considering the materials present in the labeled datasets of the first two approaches more closely, we find that carbon in a diamond-like structure is classified as a suitable candidate in the Ferrenti approach, in good agreement with experimental observations (see Table~\ref{tab:qt-materials}). Interestingly, in the labeled data from the Ferrenti approach, we find that carbon in two-dimensional graphite-like structures are marked as suitable as well. All structures of silicon are also labeled as suitable together with one entry of silicon carbide. Note that this is the $3$C polytype of SiC, meaning that the most well-established quantum compatible SiC polytype, $4$H-SiC, is missing. 
Among other potentially suitable candidates we find ZnS, ZnSe, ZnO and ZnTe present in the labeled data from the Ferrenti approach.  
In the labeled dataset of the extended Ferrenti approach, we find a single entry for each of SiC, Si, GaN, ZnS, GaP, AlAs and AlP, carbon in both diamond- and graphite-like structures, and AlN in three different configurations. The labeled dataset includes a larger variety of materials that are known to be quantum compatible as compared to the Ferrenti approach due to admitting more elements in the initial selection process. However, since we also included a more stringent band gap restriction of $1.5$~eV (to reduce the amount of materials in the training set), we find a more sparse representation of each known chemical formula present in the labeled data. 

\begin{figure}[t]
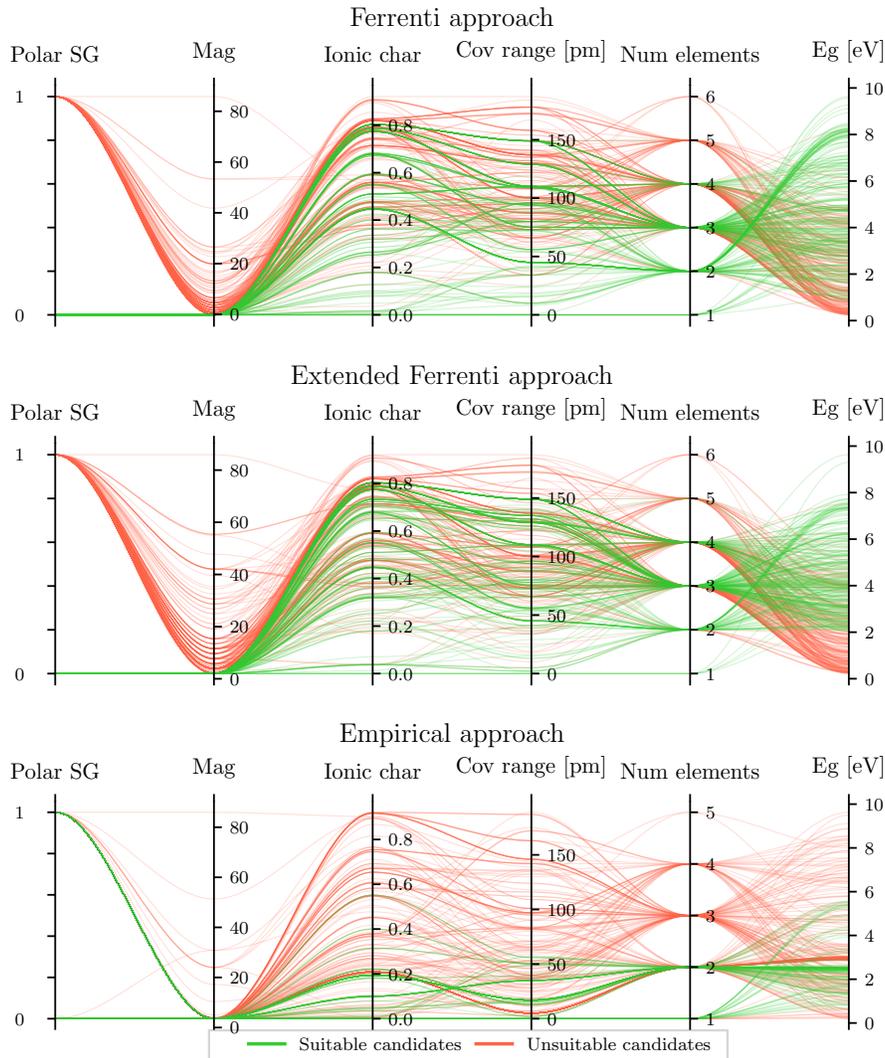
 
    \centering
    \begin{subfigure}{1\textwidth}
        \centering
        \scalebox{0.85}{\input{AllFigures/01-ferrenti-approach-v2.pgf}}
    \end{subfigure}
    \begin{subfigure}{1\textwidth}
        \centering
          \scalebox{0.85}{\input{AllFigures/02-extended-ferrenti-approach-v3.pgf}}
    \end{subfigure}
    \begin{subfigure}{1\textwidth}
        \centering
          \scalebox{0.85}{\input{AllFigures/03-empirical-approach-v3.pgf}}
    \end{subfigure}
    \caption{Parallel coordinate plots for the different approaches. To limit data cluttering, we have randomly collected up to $250$ entries for each class. For the empirical approach, we have used all the $187$ suitable candidates. The axes show total magnetization (mag), space group (SG), ionic character (ionic char), covalent range (cov range) as calculated from elemental properties, number of elements (num elements) and energy gap (Eg) as extracted from the MP database.} 
    \label{fig:parallel-coordinates-approaches}
\end{figure}

The labeled data for each approach is visualized in Fig.~\ref{fig:parallel-coordinates-approaches} as parallel coordinate plots for a few selected features informed by the criteria proposed by \citeauthor{Weber2010} \cite{Weber2010}. Parallel coordinate schemes  \cite{Inselberga1990, Inselberg1985} represent a multi-dimensional data tuple as one polyline crossing a parallel axis. The selected features are found on the $x$-axis, while the $y$-axis shows the value of the data. Thus, parallel coordinate plots can turn complex many-dimensional data into a compact  two-dimensional representation. Due to possible data cluttering, the figure visualizes a random sample of each class (suitable or unsuitable) with an upper limit of $250$ per class with transparent lines. The green and red polylines represent suitable and unsuitable candidates, respectively. 

\begin{figure}
    \centering
    \begin{subfigure}{0.35\textwidth}
        \centering
        \includegraphics[width=1\textwidth]{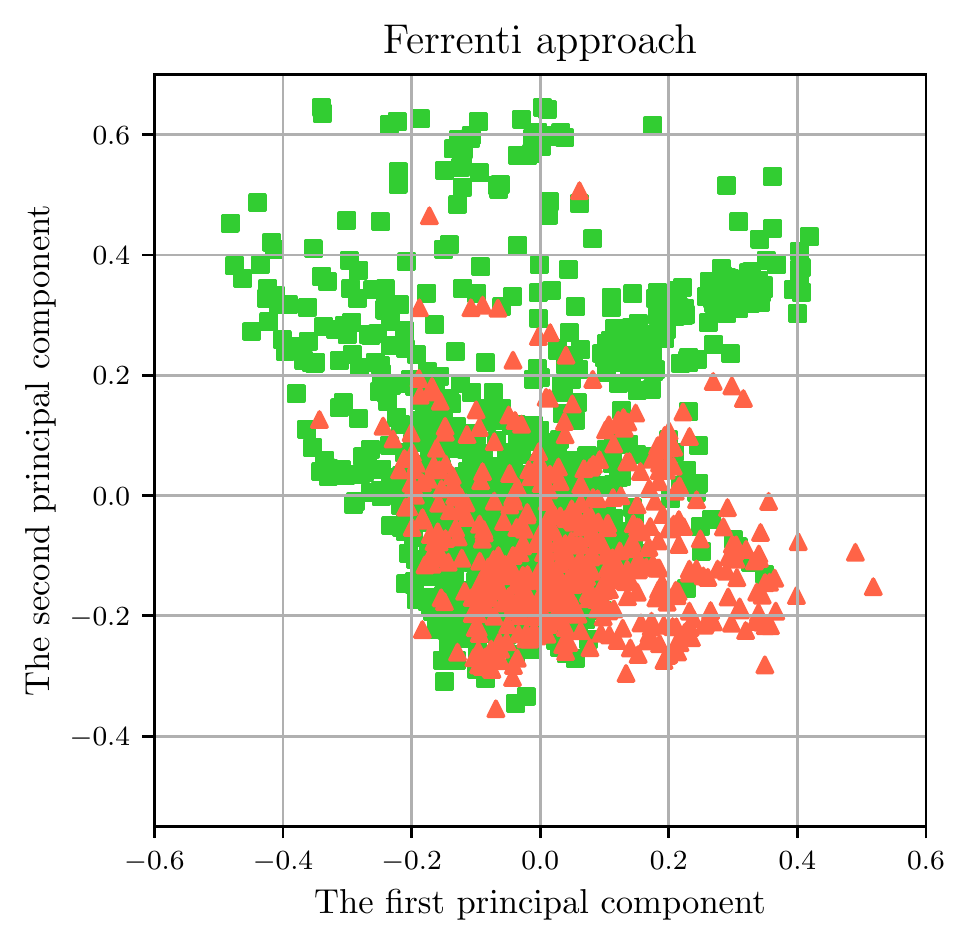}
    \end{subfigure}%
    \begin{subfigure}{0.35\textwidth}
        \centering
        \includegraphics[width=1\textwidth]{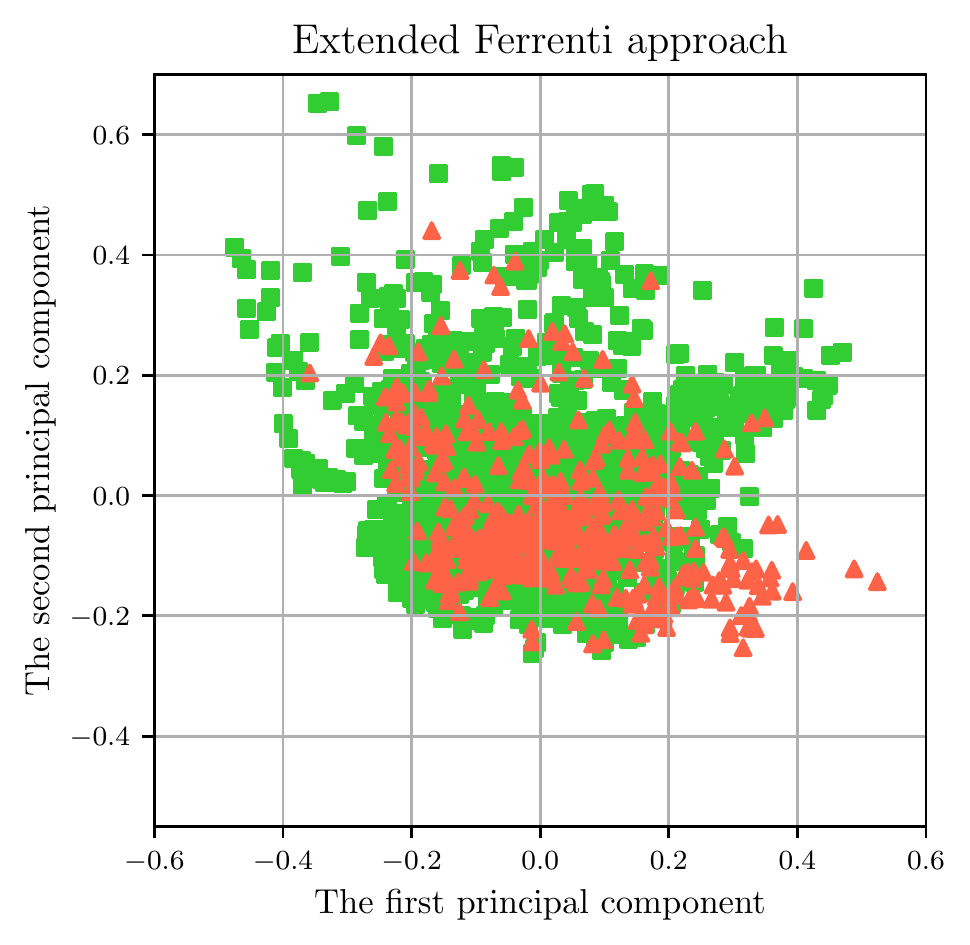}
    \end{subfigure}
    \begin{subfigure}{0.4\textwidth}
        \centering
        \includegraphics[width=1\textwidth]{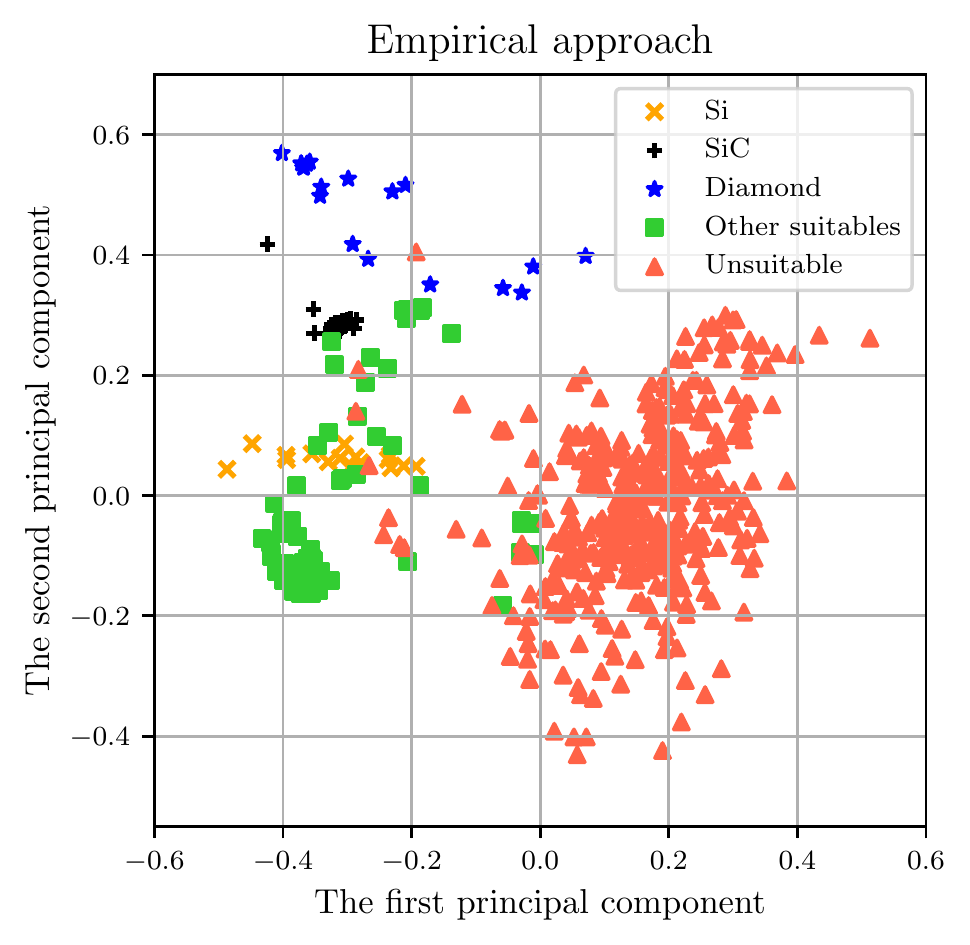}
    \end{subfigure}
    \caption{Two-dimensional scatter plots for the three different approaches. We have identified the eigenvectors corresponding to the two largest eigenvalues of the covariance-matrix, that is, the two most important principal components of the initial data from the Materials Project query. Then, we have transformed the sets of labeled data resulting from the three approaches and visualized them as scatter plots. Green squares display suitable candidates, along with the black (SiC), blue (diamond) and yellow (Si) symbols in the bottom panel, and red triangles represent unsuitable candidates.}
    \label{fig:2dscatterplotpca}
\end{figure}

The Ferrenti approach and the extended Ferrenti approach share several similarities. For example, the entries in  Fig.~\ref{fig:parallel-coordinates-approaches} span roughly the same dimensions for the Ferrenti and extended Ferrenti approaches, despite the more restricted nature of the former.  
Moreover, the two approaches' upper limits for ionic character and covalent range are very similar in the case of candidates labeled as both suitable and unsuitable.   
The most striking difference is found for the empirical approach. Again, the entries labeled as suitable candidates do not exhibit any magnetization, but now the materials crystallize in both polar and non-polar space groups --- contrarily to the two Ferrenti approaches. Moreover, we observe that the range of covalent radius and maximum ionic character span a substantially smaller parameter space in the case of the empirical approach than for the two former ones.  
Here, we can begin to deduce some trends, if the findings from the empirical approach are taken as guidance. In the empirical approach the polarity of the material's space group may be less important compared to the Ferrenti based approaches, while the accepted values for ionic character and covalent range appear more restricted (Fig.~\ref{fig:parallel-coordinates-approaches}). 

To illustrate the composition of the labeled data we perform a principal component analysis (PCA) \cite{Jolliffe2002}. Here we apply PCA to the labeled datasets in order to reveal specific features that stand out. In its standard  form, PCA relates the variance of the features with the eigenvalues of the covariance matrix \cite{Jolliffe2002,Murphy2012,Hastie2009}. We identify the two largest eigenvalues of the covariance matrix \cite{Hastie2009} of the complete initial dataset from the MP database, and transform the three labeled datasets according to the corresponding two eigenvectors. In essence, we reduce the dimensionality of the labeled data in order to visualize them.
The result of this procedure is displayed in the scatter plots of Fig.~\ref{fig:2dscatterplotpca}. Note that some minor differences between the three panels in Fig.~\ref{fig:2dscatterplotpca} may occur due to the process of removing the mean and scaling to unit variance. Red triangles represent unsuitable candidates while colored symbols (green, blue, black and yellow) represent suitable candidates. 
Due to the complexity of reducing the large amount of features down to only two, both suitable and unsuitable candidates for the Ferrenti and the extended Ferrenti approaches are largely overlapping. 
The logic behind categorizing materials in two classes (suitable and unsuitable) appears to not translate into a distinct separation, at least not in the representation of Fig.~\ref{fig:2dscatterplotpca} for the Ferrenti and extended Ferrenti approaches.  
Hence, using these approaches for predicting QT material hosts could prove challenging for any model that would try to glean a clear-cut boundary between materials that are and are not suitable for QT. 
We therefore expect that the two Ferrenti approaches could need supplementary dimensions for further distinguishing between the materials in the two categories (suitable and unsuitable), or that it may prove challenging to find a generalized model based on the training sets that were defined according to the above discussion. 

In the case of the empirical approach, on the other hand, we can discern a trend where the upper left part of Fig.~\ref{fig:2dscatterplotpca} is dominated by suitable candidates (green, blue, yellow and black symbols), while the unsuitable candidates (red triangles) are similarly restricted to the lower right corner, albeit with some exceptions. 
Interestingly, we observe that different configurations of the famously quantum compatible materials of diamond (blue), silicon carbide (black) and silicon (yellow) are grouped together but each material class is separated in its own region. 
This indicates that the empirical approach may be capable of separating materials based on their underlying properties. Thus, devising a logical framework for the data mining process is likely important in order to obtain suitable datasets for training and testing of the ML methods. 


\subsection*{Machine learning and principal component analysis}
We have applied four well-tested ML algorithms in order to classify specific materials as candidate systems for QT. These are logistic regression, decision trees, and the ensemble methods random forests and gradient boosting \cite{Mehta2019,Hastie2009,Murphy2012}. 
These four ML methods were applied to the labeled data sets that were optimized using each of the three approaches (Ferrenti, extended Ferrenti and empirical) outlined above, yielding twelve sets of candidates that were predicted as suitable material hosts for QT applications.  

In order to reduce the dimensionality while preserving most of the variance in the data, 
we have performed a principal component analysis  \cite{Jolliffe2002} for each of the three approaches (Ferrenti, extended Ferrenti and empirical).
First, the eigenvalues of the covariance matrices of the three training sets are identified in order of decreasing value, resulting in three distinct PCA models (as opposed to the case of Fig.~\ref{fig:2dscatterplotpca} where all panels are based on the same model). 
Next, in the evaluation of the ML methods for the three different approaches we apply a $5\times 5$ stratified cross-validation \cite{Hastie2009} when searching for the optimal hyperparameter combinations. We note that all four ML methods have high evaluation metrics for the optimal hyperparameters. Further details on the evaluation of the ML methods and the results from the principal component analyses for all four methods are shown in the Supplementary Information \cite{supplementary}.  
The three data labeling approaches (Ferrenti, extended Ferrenti and empirical) define sets of labeled data of varying sizes, where the smallest is from the empirical approach. For small datasets, it is proven to be beneficial to repeat the cross-validation analysis \cite{Hastie2009}, as this is a method that allows us to measure the stability of the predictions against perturbations (i.e., few different entries) in the training data \cite{Beleites2008}.

\begin{figure}[t]
    \centering
    \begin{subfigure}[b]{0.45\textwidth}
\begin{tikzpicture}

\definecolor{color0}{rgb}{0.8,0.4,0.466666666666667}

\begin{axis}[
height=1.6222438079424382in,
legend style={fill opacity=0.8, draw opacity=1, text opacity=1, draw=white!80!black},
tick align=outside,
tick pos=left,
width=3.0444876158848764in,
x grid style={white!69.0196078431373!black},
xmajorgrids,
xmin=0.5, xmax=26,
xtick style={color=black},
y grid style={white!69.0196078431373!black},
ymajorgrids,
ymin=0, ymax=0.133808014644924,
ytick style={color=black},
yticklabels={, 0, , 0.1},
title={Explained variance},
legend style={font=\footnotesize},
]
\draw[draw=none,fill=white!53.3333333333333!black,fill opacity=0.5] (axis cs:0.6,0) rectangle (axis cs:1.4,0.123808014644924);
\draw[draw=none,fill=white!53.3333333333333!black,fill opacity=0.5] (axis cs:1.6,0) rectangle (axis cs:2.4,0.0779020383908885);
\draw[draw=none,fill=white!53.3333333333333!black,fill opacity=0.5] (axis cs:2.6,0) rectangle (axis cs:3.4,0.0690715494994265);
\draw[draw=none,fill=white!53.3333333333333!black,fill opacity=0.5] (axis cs:3.6,0) rectangle (axis cs:4.4,0.0448634890553009);
\draw[draw=none,fill=white!53.3333333333333!black,fill opacity=0.5] (axis cs:4.6,0) rectangle (axis cs:5.4,0.0396017649392185);
\draw[draw=none,fill=white!53.3333333333333!black,fill opacity=0.5] (axis cs:5.6,0) rectangle (axis cs:6.4,0.0309892340254588);
\draw[draw=none,fill=white!53.3333333333333!black,fill opacity=0.5] (axis cs:6.6,0) rectangle (axis cs:7.4,0.0286471759268761);
\draw[draw=none,fill=white!53.3333333333333!black,fill opacity=0.5] (axis cs:7.6,0) rectangle (axis cs:8.4,0.0241373022722755);
\draw[draw=none,fill=white!53.3333333333333!black,fill opacity=0.5] (axis cs:8.6,0) rectangle (axis cs:9.4,0.0219387724823991);
\draw[draw=none,fill=white!53.3333333333333!black,fill opacity=0.5] (axis cs:9.6,0) rectangle (axis cs:10.4,0.0191941500093896);
\draw[draw=none,fill=white!53.3333333333333!black,fill opacity=0.5] (axis cs:10.6,0) rectangle (axis cs:11.4,0.0170987447716404);
\draw[draw=none,fill=white!53.3333333333333!black,fill opacity=0.5] (axis cs:11.6,0) rectangle (axis cs:12.4,0.0163324377943715);
\draw[draw=none,fill=white!53.3333333333333!black,fill opacity=0.5] (axis cs:12.6,0) rectangle (axis cs:13.4,0.0142479118996461);
\draw[draw=none,fill=white!53.3333333333333!black,fill opacity=0.5] (axis cs:13.6,0) rectangle (axis cs:14.4,0.0139471831901937);
\draw[draw=none,fill=white!53.3333333333333!black,fill opacity=0.5] (axis cs:14.6,0) rectangle (axis cs:15.4,0.0125561362477749);
\draw[draw=none,fill=white!53.3333333333333!black,fill opacity=0.5] (axis cs:15.6,0) rectangle (axis cs:16.4,0.0122752355665442);
\draw[draw=none,fill=white!53.3333333333333!black,fill opacity=0.5] (axis cs:16.6,0) rectangle (axis cs:17.4,0.0117377683648439);
\draw[draw=none,fill=white!53.3333333333333!black,fill opacity=0.5] (axis cs:17.6,0) rectangle (axis cs:18.4,0.0108124054147357);
\draw[draw=none,fill=white!53.3333333333333!black,fill opacity=0.5] (axis cs:18.6,0) rectangle (axis cs:19.4,0.0105912984624588);
\draw[draw=none,fill=white!53.3333333333333!black,fill opacity=0.5] (axis cs:19.6,0) rectangle (axis cs:20.4,0.00987028371434172);
\draw[draw=none,fill=white!53.3333333333333!black,fill opacity=0.5] (axis cs:20.6,0) rectangle (axis cs:21.4,0.00946157777094173);
\draw[draw=none,fill=white!53.3333333333333!black,fill opacity=0.5] (axis cs:21.6,0) rectangle (axis cs:22.4,0.00932051055389097);
\draw[draw=none,fill=white!53.3333333333333!black,fill opacity=0.5] (axis cs:22.6,0) rectangle (axis cs:23.4,0.00851928669553865);
\draw[draw=none,fill=white!53.3333333333333!black,fill opacity=0.5] (axis cs:23.6,0) rectangle (axis cs:24.4,0.00821174057436937);
\draw[draw=none,fill=white!53.3333333333333!black,fill opacity=0.5] (axis cs:24.6,0) rectangle (axis cs:25.4,0.00796122578616887);
\addlegendimage{semithick, color=white!53.3333333333333!black};
\addlegendentry{Ferrenti approach};
\end{axis}

\end{tikzpicture}
        \label{fig:01-fi-e}
    \end{subfigure}
    \begin{subfigure}[b]{0.45\textwidth}
\begin{tikzpicture}

\definecolor{color0}{rgb}{0.0666666666666667,0.466666666666667,0.2}

\begin{axis}[
height=1.6222438079424382in,
tick align=outside,
tick pos=left,
width=3.0444876158848764in,
x grid style={white!69.0196078431373!black},
xmajorgrids,
xmin=0.5, xmax=26,
xtick style={color=black},
y grid style={white!69.0196078431373!black},
ymajorgrids,
ymin=0, ymax=0.241504672452706,
ytick style={color=black},
yticklabels={,0, 0.1,0.2},
title={Gradient boosting feature importance},
legend style={font=\footnotesize},
]
\draw[draw=none,fill=color0] (axis cs:0.6,0) rectangle (axis cs:1.4,0.0939219294354606);
\draw[draw=none,fill=color0] (axis cs:1.6,0) rectangle (axis cs:2.4,0.128717126999015);
\draw[draw=none,fill=color0] (axis cs:2.6,0) rectangle (axis cs:3.4,0.0213938886864563);
\draw[draw=none,fill=color0] (axis cs:3.6,0) rectangle (axis cs:4.4,0.00888144685168664);
\draw[draw=none,fill=color0] (axis cs:4.6,0) rectangle (axis cs:5.4,0.191504672452706);
\draw[draw=none,fill=color0] (axis cs:5.6,0) rectangle (axis cs:6.4,0.00724897098092836);
\draw[draw=none,fill=color0] (axis cs:6.6,0) rectangle (axis cs:7.4,0.00450287540874152);
\draw[draw=none,fill=color0] (axis cs:7.6,0) rectangle (axis cs:8.4,0.00389892865504545);
\draw[draw=none,fill=color0] (axis cs:8.6,0) rectangle (axis cs:9.4,0.00328938623970718);
\draw[draw=none,fill=color0] (axis cs:9.6,0) rectangle (axis cs:10.4,0.00607634168998537);
\draw[draw=none,fill=color0] (axis cs:10.6,0) rectangle (axis cs:11.4,0.0075927359484662);
\draw[draw=none,fill=color0] (axis cs:11.6,0) rectangle (axis cs:12.4,0.00246964434307127);
\draw[draw=none,fill=color0] (axis cs:12.6,0) rectangle (axis cs:13.4,0.0113273434472604);
\draw[draw=none,fill=color0] (axis cs:13.6,0) rectangle (axis cs:14.4,0.0215526621032538);
\draw[draw=none,fill=color0] (axis cs:14.6,0) rectangle (axis cs:15.4,0.00889067969625317);
\draw[draw=none,fill=color0] (axis cs:15.6,0) rectangle (axis cs:16.4,0.0112833024636162);
\draw[draw=none,fill=color0] (axis cs:16.6,0) rectangle (axis cs:17.4,0.0331145663678211);
\draw[draw=none,fill=color0] (axis cs:17.6,0) rectangle (axis cs:18.4,0.0265656605925389);
\draw[draw=none,fill=color0] (axis cs:18.6,0) rectangle (axis cs:19.4,0.0412663962030667);
\draw[draw=none,fill=color0] (axis cs:19.6,0) rectangle (axis cs:20.4,0.00893277229511404);
\draw[draw=none,fill=color0] (axis cs:20.6,0) rectangle (axis cs:21.4,0.00638009737027819);
\draw[draw=none,fill=color0] (axis cs:21.6,0) rectangle (axis cs:22.4,0.0110083486001703);
\draw[draw=none,fill=color0] (axis cs:22.6,0) rectangle (axis cs:23.4,0.00396903870317716);
\draw[draw=none,fill=color0] (axis cs:23.6,0) rectangle (axis cs:24.4,0.0182603050195842);
\draw[draw=none,fill=color0] (axis cs:24.6,0) rectangle (axis cs:25.4,0.0400563256690563);
\path [draw=color0, semithick]
(axis cs:1,0.0728244166732798)
--(axis cs:1,0.115019442197641);

\path [draw=color0, semithick]
(axis cs:2,0.102746647015722)
--(axis cs:2,0.154687606982308);

\path [draw=color0, semithick]
(axis cs:3,0.0096504157614985)
--(axis cs:3,0.0331373616114142);

\path [draw=color0, semithick]
(axis cs:4,0.00120429954877644)
--(axis cs:4,0.0165585941545968);

\path [draw=color0, semithick]
(axis cs:5,0.164674846558901)
--(axis cs:5,0.218334498346511);

\path [draw=color0, semithick]
(axis cs:6,0.00131865659765555)
--(axis cs:6,0.0131792853642012);

\path [draw=color0, semithick]
(axis cs:7,0.00089826602034204)
--(axis cs:7,0.008107484797141);

\path [draw=color0, semithick]
(axis cs:8,-0.000515167535189504)
--(axis cs:8,0.00831302484528041);

\path [draw=color0, semithick]
(axis cs:9,0.000577548695504199)
--(axis cs:9,0.00600122378391017);

\path [draw=color0, semithick]
(axis cs:10,-0.00123032392384619)
--(axis cs:10,0.0133830073038169);

\path [draw=color0, semithick]
(axis cs:11,0.00341822410793035)
--(axis cs:11,0.0117672477890021);

\path [draw=color0, semithick]
(axis cs:12,0.000427923119383559)
--(axis cs:12,0.00451136556675897);

\path [draw=color0, semithick]
(axis cs:13,-0.000894751232806041)
--(axis cs:13,0.0235494381273268);

\path [draw=color0, semithick]
(axis cs:14,-0.00534323462302397)
--(axis cs:14,0.0484485588295317);

\path [draw=color0, semithick]
(axis cs:15,-0.00580574316011168)
--(axis cs:15,0.023587102552618);

\path [draw=color0, semithick]
(axis cs:16,-0.0063295986999169)
--(axis cs:16,0.0288962036271493);

\path [draw=color0, semithick]
(axis cs:17,-0.000960237853350046)
--(axis cs:17,0.0671893705889922);

\path [draw=color0, semithick]
(axis cs:18,-0.00343303381845922)
--(axis cs:18,0.056564355003537);

\path [draw=color0, semithick]
(axis cs:19,0.00206699809958492)
--(axis cs:19,0.0804657943065485);

\path [draw=color0, semithick]
(axis cs:20,-0.00625580047143954)
--(axis cs:20,0.0241213450616676);

\path [draw=color0, semithick]
(axis cs:21,-0.00190515682114497)
--(axis cs:21,0.0146653515617014);

\path [draw=color0, semithick]
(axis cs:22,-0.00451438685233842)
--(axis cs:22,0.0265310840526791);

\path [draw=color0, semithick]
(axis cs:23,-0.000399848870586143)
--(axis cs:23,0.00833792627694046);

\path [draw=color0, semithick]
(axis cs:24,-0.0070765748506951)
--(axis cs:24,0.0435971848898635);

\path [draw=color0, semithick]
(axis cs:25,-0.00342720675075281)
--(axis cs:25,0.0835398580888655);
\addlegendimage{semithick, color=color0};
\addlegendentry{Ferrenti approach};
\addplot [semithick, color0, mark=-, mark size=0, mark options={solid}, only marks]
table {%
1 0.0728244166732798
2 0.102746647015722
3 0.0096504157614985
4 0.00120429954877644
5 0.164674846558901
6 0.00131865659765555
7 0.00089826602034204
8 -0.000515167535189504
9 0.000577548695504199
10 -0.00123032392384619
11 0.00341822410793035
12 0.000427923119383559
13 -0.000894751232806041
14 -0.00534323462302397
15 -0.00580574316011168
16 -0.0063295986999169
17 -0.000960237853350046
18 -0.00343303381845922
19 0.00206699809958492
20 -0.00625580047143954
21 -0.00190515682114497
22 -0.00451438685233842
23 -0.000399848870586143
24 -0.0070765748506951
25 -0.00342720675075281
};
\addplot [semithick, color0, mark=-, mark size=4, mark options={solid}, only marks]
table {%
1 0.115019442197641
2 0.154687606982308
3 0.0331373616114142
4 0.0165585941545968
5 0.218334498346511
6 0.0131792853642012
7 0.008107484797141
8 0.00831302484528041
9 0.00600122378391017
10 0.0133830073038169
11 0.0117672477890021
12 0.00451136556675897
13 0.0235494381273268
14 0.0484485588295317
15 0.023587102552618
16 0.0288962036271493
17 0.0671893705889922
18 0.056564355003537
19 0.0804657943065485
20 0.0241213450616676
21 0.0146653515617014
22 0.0265310840526791
23 0.00833792627694046
24 0.0435971848898635
25 0.0835398580888655
};
\end{axis}

\end{tikzpicture}
        \label{fig:01-fi-d}
    \end{subfigure}%
    \hfill
    \begin{subfigure}[b]{0.45\textwidth}
        \input{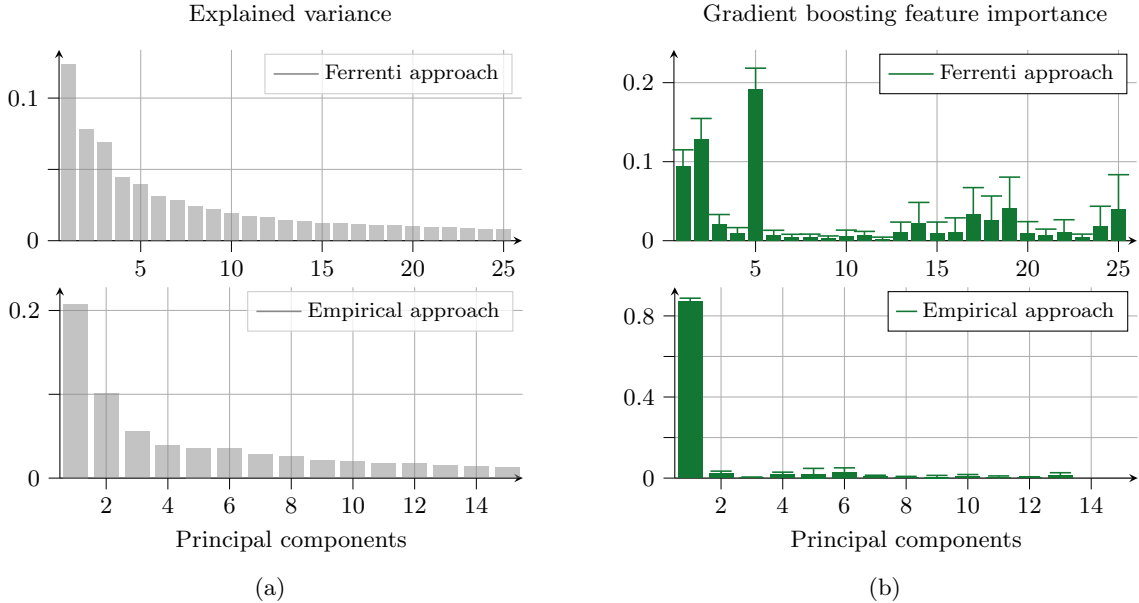}
        \label{fig:03-fi-e}
        \subcaption{}
    \end{subfigure}
    \begin{subfigure}[b]{0.45\textwidth}
\begin{tikzpicture}

\definecolor{color0}{rgb}{0.0666666666666667,0.466666666666667,0.2}

\begin{axis}[
height=1.6222438079424382in,
tick align=outside,
tick pos=left,
width=3.0444876158848764in,
x grid style={white!69.0196078431373!black},
xmajorgrids,
xmin=0.5, xmax=15.5,
xlabel={Principal components},
xtick style={color=black},
y grid style={white!69.0196078431373!black},
ymajorgrids,
ymin=0, ymax=0.941178662009005,
yticklabels={,0,, 0.4,,0.8},
ytick style={color=black},
legend style={font=\footnotesize}
]
\draw[draw=none,fill=color0] (axis cs:0.6,0) rectangle (axis cs:1.4,0.871178662009005);
\draw[draw=none,fill=color0] (axis cs:1.6,0) rectangle (axis cs:2.4,0.0217568511840037);
\draw[draw=none,fill=color0] (axis cs:2.6,0) rectangle (axis cs:3.4,0.00253604996390695);
\draw[draw=none,fill=color0] (axis cs:3.6,0) rectangle (axis cs:4.4,0.0192708196133984);
\draw[draw=none,fill=color0] (axis cs:4.6,0) rectangle (axis cs:5.4,0.0198922275506549);
\draw[draw=none,fill=color0] (axis cs:5.6,0) rectangle (axis cs:6.4,0.0260375424696281);
\draw[draw=none,fill=color0] (axis cs:6.6,0) rectangle (axis cs:7.4,0.0063356815006726);
\draw[draw=none,fill=color0] (axis cs:7.6,0) rectangle (axis cs:8.4,0.00319869517406271);
\draw[draw=none,fill=color0] (axis cs:8.6,0) rectangle (axis cs:9.4,0.00540195176007804);
\draw[draw=none,fill=color0] (axis cs:9.6,0) rectangle (axis cs:10.4,0.00695771640251456);
\draw[draw=none,fill=color0] (axis cs:10.6,0) rectangle (axis cs:11.4,0.00312270971606427);
\draw[draw=none,fill=color0] (axis cs:11.6,0) rectangle (axis cs:12.4,0.0030932299465922);
\draw[draw=none,fill=color0] (axis cs:12.6,0) rectangle (axis cs:13.4,0.0112178627094187);
\path [draw=color0, semithick]
(axis cs:1,0.854956014167526)
--(axis cs:1,0.887401309850484);

\path [draw=color0, semithick]
(axis cs:2,0.00907965200713417)
--(axis cs:2,0.0344340503608733);

\path [draw=color0, semithick]
(axis cs:3,-0.000182654889842383)
--(axis cs:3,0.00525475481765629);

\path [draw=color0, semithick]
(axis cs:4,0.00923661909617688)
--(axis cs:4,0.0293050201306199);

\path [draw=color0, semithick]
(axis cs:5,-0.00820788248180509)
--(axis cs:5,0.0479923375831149);

\path [draw=color0, semithick]
(axis cs:6,0.00143690786973603)
--(axis cs:6,0.0506381770695201);

\path [draw=color0, semithick]
(axis cs:7,-0.00149882818592621)
--(axis cs:7,0.0141701911872714);

\path [draw=color0, semithick]
(axis cs:8,-0.00213482608794717)
--(axis cs:8,0.00853221643607259);

\path [draw=color0, semithick]
(axis cs:9,-0.00260933890414728)
--(axis cs:9,0.0134132424243034);

\path [draw=color0, semithick]
(axis cs:10,-0.0038254976017747)
--(axis cs:10,0.0177409304068038);

\path [draw=color0, semithick]
(axis cs:11,-0.0044786262895489)
--(axis cs:11,0.0107240457216774);

\path [draw=color0, semithick]
(axis cs:12,-0.000395131606741384)
--(axis cs:12,0.00658159149992578);

\path [draw=color0, semithick]
(axis cs:13,-0.00413997684116445)
--(axis cs:13,0.0265757022600018);

\addplot [semithick, color0, mark=-, mark size=4, mark options={solid}, only marks]
table {%
1 0.854956014167526
2 0.00907965200713417
3 -0.000182654889842383
4 0.00923661909617688
5 -0.00820788248180509
6 0.00143690786973603
7 -0.00149882818592621
8 -0.00213482608794717
9 -0.00260933890414728
10 -0.0038254976017747
11 -0.0044786262895489
12 -0.000395131606741384
13 -0.00413997684116445
};
\addplot [semithick, color0, mark=-, mark size=4, mark options={solid}, only marks]
table {%
1 0.887401309850484
2 0.0344340503608733
3 0.00525475481765629
4 0.0293050201306199
5 0.0479923375831149
6 0.0506381770695201
7 0.0141701911872714
8 0.00853221643607259
9 0.0134132424243034
10 0.0177409304068038
11 0.0107240457216774
12 0.00658159149992578
13 0.0265757022600018
};
\addlegendimage{semithick, color=color0};
\addlegendentry{Empirical approach};
\end{axis}

\end{tikzpicture}
        \label{fig:03-fi-d}
        \subcaption{}
    \end{subfigure}
    \caption{Visualization of different parameters for the $25$ most principal components ranked in descending order by (a) the explained variance for the Ferrenti (upper panel) and empirical (lower panel) approaches, and (b) the gradient boosting coefficients for the corresponding approaches. Note that in (a) we visualize only the results of the PCA analysis while in (b) there are results of the PCA reduced training sets using gradient boosting \cite{Hastie2009,xgboost2016}. The latter shows that for the empirical approach most of the physics is represented by a few features. The results in (b) are similar to those obtained with logistic regression, decision trees and random forests.  
    }
    \label{fig:PComponents}
\end{figure}

Figure~\ref{fig:PComponents} visualizes different parameters for the most important principal components ranked in descending order by (a) the explained variance for the Ferrenti (upper panel) and empirical (lower panel) approach, and (b) the gradient boosting coefficients for the corresponding approaches. This differs from Fig.~\ref{fig:2dscatterplotpca}, where we visualize the two most important eigenvectors for all approaches that originates from the same covariance matrix. Note also that in  Fig.~\ref{fig:2dscatterplotpca} we performed a PCA analysis on all data. Here, the ML analysis is done on the labeled data for each approach. For the Ferrenti approach (Fig.~\ref{fig:PComponents}a) to reach the $95 \ \%$ accumulated explained variance, a total of $144$ principal components must be included. For the extended Ferrenti approach, we find that gradient boosting performs optimally at $93$ principal components. 
In the case of the empirical approach and in contrast to the Ferrenti approaches, decision trees and random forests exhibit the best performance for just a few principal components, and experience a considerable degree of overfitting when involving more principal components. Gradient boosting 
also experiences the best performance for just a few principal components. 

Figure~\ref{fig:PComponents}(b) reveals that the most important feature for the gradient boosting method in the Ferrenti approach is the fifth principal component (fifth largest eigenvalue). The trend is maintained for all four ML methods (see the Supplementary Information at \cite{supplementary}). Moreover, by selecting the highest values in this eigenvector, we find that the corresponding features originate from the DFT computed band gap of the elemental solids among the elements in the compound as calculated by the Materials Agnostic Platform for Informatics and Exploration (MagPie)~\cite{magpie}. 
The second most important principal component for the Ferrenti approach exhibits significant contributions from the covalent radius, the ionic character and the packing efficiency among the elements in the composition. 
The data originates from elemental calculations from MagPie and are aggregated as either minimum, mean, standard deviation, or maximum. 
While the first principal component encompasses the largest explained variance, it does not provide a clear and specific information on which features it represents. 

The interpretation of feature importance for the extended Ferrenti approach is substantially more challenging than in the Ferrenti approach. We find for logistic regression and decision trees that no feature is different than any other in the cross-validation, mainly due to a large variety of accuracy. However, we find that random forests and gradient boosting both mark the fifth principal component as important. Similar to the Ferrenti approach, the corresponding features with the highest value for the fifth principal component originate from the DFT computed band gap of the elemental solids among the elements in the composition. 

The empirical approach differs in many aspects from the Ferrenti and extended Ferrenti approaches. 
Firstly, we find that the number of principal components necessary to obtain $95 \ \%$ variance is reduced to $103$ components, which is $41$ and $56$ less than the Ferrenti and extended Ferrenti approaches, respectively. Thus, the variance of the training set is found to be described with fewer principal components, indicating a simpler model. Secondly, the first principal component is by far the most important feature for all ML methods in the empirical approach, as visualized in the lower panel of Fig.~\ref{fig:PComponents}(b) for gradient boosting. Similar conclusions are reached when using logistic regression, decisions trees and random forests as classification methods (see the Supplementary Information at \cite{supplementary}). 
The distinct importance of the first principal component partly explains why we experience a high accuracy for only a single feature. The first principal component's corresponding features is a complex combination of several material properties, but we find that it includes bond orientational parameters, coordination numbers, and the radial distribution function of a compound's crystal system. 
The standard deviation of the radial distribution function appears multiple times in the list of features and is of particular importance. 
Thirdly, the empirical approach differs in how much explained variance is retained by the first component, which is $21 \ \%$, while it is $14 \ \%$ for the Ferrenti approach and $11 \ \%$ for the extended Ferrenti approach. We find the difference striking considering that the approaches share the same ultimate goal, but where the training sets apparently exhibit large and significant variations. 

Intriguingly, the principal components that are deemed important by the three approaches differ substantially. The Ferrenti and extended Ferrenti approaches place particular value on the material's band gap and the ionic or covalent character of the bonding. Indeed, precisely these features were used as guidance in the data labeling process. Thus, the ML methods are preserving the criteria imposed in the Ferrenti approaches and perpetuate our initial assumptions on which material properties to look for. For the empirical approach, on the other hand, we did not guide the selection process in terms of specific properties. Here, the ML methods appear to be informed by other  characteristics than band gap and ionic character that are more related to symmetry and crystal structure, based on the repeated appearance of bond length, orientation and radial distribution in the first principal component. In the empirical approach, the ML methods appear to be inferring that the common trend among the materials that are known to be suitable for QT is related to more complex mechanisms in the crystal structure and bonding. 

\subsection*{Predicting Suitable Material Hosts for Quantum Technology} 
After training the ML algorithms using parts of the data labeled during the three data mining approaches, the ML methods were employed on the test sets. 
The number of candidates predicted by each of the four ML methods logistic regression, decision trees, random forests and gradient boosting is visualized in the Supplementary Information at \cite{supplementary} for each of the three approaches (Ferrenti, extended Ferrenti and empirical). 
An overview of the number of predicted suitable candidates as a function of a given ML algorithm's confidence is summarized in  Table~\ref{tab:probabilites}. 

\subsubsection*{The Ferrenti approach}

From the predicted dataset of $23.623$ materials 
the four ML methods consistently predict at least $11.000$ materials as promising candidates in the case of the Ferrenti approach. All four ML methods agree on a total of $6804$ suitable candidates, however, many of these materials are predicted with an incidence similar to that of a coin flip. If we were to raise the minimum confidence cut-off for a prediction to, e.g., $0.75$ instead of $0.5$, the above ML algorithms would only agree on $1784$ suitable candidates. 
We find that all four ML methods admit almost all materials with a chemical formula matching the known suitable candidates (see Table~\ref{tab:qt-materials}) that were present in the labeled data. This can allow materials with unfortunate structures to be labeled as suitable candidates. Notably, the ML methods do not maintain the band gap restriction from the training set definition, where all materials with band gaps lower than $0.5$~eV were eliminated. 
This trend is not carried over to the predicted data. Indeed, we find that many entries with band gaps lower than $0.5$~eV are predicted as suitable candidates by all four ML methods employed herein 
--- despite the principal component analysis revealing that the band gap is an important feature for the machine learning classification in the Ferrenti approach.  
On the other hand, due to the known  underestimation of band gaps by the PBE  functional, the band gaps could in reality be larger for many of the materials. 

\begin{table}[t]
    \centering 
    \caption{Overview of the number of predicted suitable candidates as a function of a given ML method's confidence when the four methods in an approach agree. A threshold value of $0.5$ represents the confidence of a coin flip while $1.0$ is a fully confident prediction.}
    \begin{tabular}{c|c|c|c}
      & $0.50$ & $0.75$ & $0.85$ \\
     \hline
     Ferrenti approach &  $6804$ & $1784$ & $258$  \\
     Extended Ferrenti approach &  $9227$ & $4735$  & $2001$  \\ 
     Empirical approach & $214$ & $66$ & $9$ \\
     \hline
     All approaches together & $47$ & $6$ & 0 \\
    \end{tabular}
    \label{tab:probabilites}
\end{table}

\subsubsection*{The extended Ferrenti approach}
Next, we turn to the more liberal extended Ferrenti approach containing a $78 \ \%$ larger 
amount of labeled data than the Ferrenti approach.  
The four ML methods predict at least $13.000$ materials as suitable candidates out of $22.550$ materials in the unlabeled dataset, where they all agree on a total of $9227$ predicted suitable candidates. Comparing to the labeled data reveals that all of the unlabeled candidates that are known to be suitable (see Table~\ref{tab:qt-materials}) are, in fact, predicted as suitable candidates. 
However, the ML methods also predict materials as suitable that we would not expect according to, e.g., \citeauthor{Weber2010} \cite{Weber2010}. 
All four ML algorithms predict NaCl as a suitable candidate to confidences of $0.83$ and $0.60$ for two different configurations, despite the strong electrostatic interactions between Na and Cl and the ionic character of their bonding.  
Furthermore, although we enforced a conservative band gap restriction of $1.5$~eV for the data labeling in the extended Ferrenti approach, we still find that all four implemented ML methods predict suitable candidates that exhibit band gaps substantially lower than $0.5$~eV.  
As seen above, the ML methods applied to data labeled in the Ferrenti and extended Ferrenti approach are recognizing the band gap and bonding character as important, but the resulting predicted materials do not strictly follow the anticipated guidelines. 
 
\subsubsection*{The empirical approach}
The ML methods that were trained on the data extracted in the empirical approach predict substantially fewer candidates as compared to the other two approaches. 
A total of $842$, $1197$, $543$ and $596$ materials are classified as suitable candidates by logistic regression, decision trees, random forests and gradient boosting, respectively. All the four ML methods agree on a total of $214$ suitable candidates to $0.5$ confidence.   
Note that $51$ of these have a band gap of $0.5$~eV or smaller as reported in Materials Project \cite{Jain2013,Jain2018} from PBE-level DFT calculations. Increasing the threshold to $0.75$ or $0.85$ yields $66$ or $9$ predicted suitable candidate materials in the empirical approach, respectively (see~Table~\ref{tab:probabilites}). 

\begin{figure}[t]
    \centering
    \includegraphics[width=0.85\textwidth]{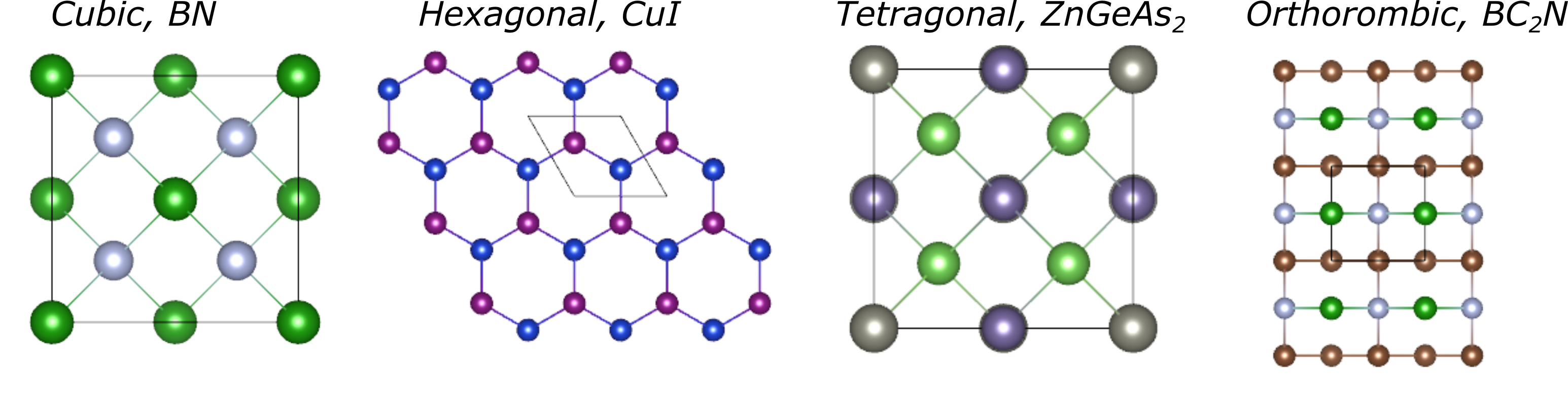}
    \caption{Illustrations of the crystal systems that the $9$ materials predicted by the empirical approach to $0.85$ confidence, and the $6$ materials predicted by all approaches to $0.75$ confidence, may belong to. Four different symmetry categories are observed; cubic, hexagonal, tetragonal and orthorombic. The viewpoints for all materials are down along the $c$-axis.   }
    \label{fig:crystalsystems}
\end{figure}

Consider the $9$ materials that were classified as suitable to a confidence of $0.85$ or higher by all four ML methods in the empirical approach; BN, CdSe (2 structures), BC$_2$N (2 structures), InAs, CuI (2 structures), and ZnCd$_3$Se$_4$. 
The nine materials (considering different crystal structures) each belong to one of the four crystal systems cubic, hexagonal, tetragonal and orthorombic. Figure~\ref{fig:crystalsystems} visualizes the four different crystal systems while Table~\ref{tab:materialproperties} lists important material properties of the relevant materials as reported in the MP database.  Interestingly, we find that all nine materials appear to be four-fold coordinated, and that the first two lattice vectors ($a$ and $b$) are identical in all nine cases whereas $c$ may differ.  
There appears to be a high degree of symmetry present for all materials, but we find no elemental and perfectly symmetric semiconductors in the list. 

Turning to each of the individual material predictions, the compound BN (mp-$1639$) was in fact present in the training data as a suitable candidate, and we believe this to be a partial explanation for why the ML methods recognize BN as a suitable candidate with high probability. Furthermore, two compositions of CdSe (the cubic mp-$2691$ and the hexagonal mp-$1070$) have been predicted as suitable, possibly as a consequence of the similar compound CdS being labeled as a suitable candidate in the training set. 
The two compounds of CdSe share similar properties with calculated band gaps in the MP database of $0.5$~eV and $0.6$~eV, respectively. 

In the case of BC$_2$N we find two compositions with the same chemical formula; the orthorhombically coordinated (mp-$629458$) and the tetragonally structured (mp-$1008523$) BC$_2$N. The former takes on a polar space group while the latter does not. The band gaps are listed as $1.9$~eV and $1.7$~eV, respectively, in the MP database. Both structures exhibit, as expected, a strong covalent character and have been studied for applications as nanostructures for electronic devices \cite{Gao2017}, hydrogen storage \cite{Cai2017} and superhard materials \cite{Li2017, Jiang2020}. Importantly, the diamond-like structure of BC$_3$N was recently predicted as a promising  spin qubit material host \cite{Wang2020SpinQB}. By creating a boron vacancy one can in theory obtain a defect center with similar properties to those found for the NV center in diamond. Whether this is also possible for BC$_2$N remains to be seen. We note that BC$_3$N was not represented in the MP database at the time of data extraction and is therefore not included in our dataset. 

The compounds InAs (cubic, mp-$20305$), CuI (cubic, mp-$22895$ and hexagonal, mp-$569346$) and ZnCd$_3$Se$_4$ (cubic, mp-$1078597$) are listed in the MP database with band gaps of $0.3$, $1.2$, $1.2$ and $1.7$~eV, respectively. 
To the best of our knowledge, ZnCd$_3$Se$_4$ has yet to be synthesized, while self-assembled InAs quantum dots are  exciting possible materials to use in quantum technology \cite{Liu2018}. 
CuI has recently been synthesized as single-crystal epitaxial films and was shown to exhibit remarkable optoelectronic properties \cite{Ahn2020}. Interestingly, the material exhibits a large ionic character with more similarities to some of the oxide compounds in the dataset than, e.g., the materials that are more covalent in character like Si, SiC and  diamond.  
The prediction of CuI in two configurations indicates that ionic character alone is not an obstacle for a material to be quantum compatible. It is unknown at this time whether any potentially favorable properties of CuI would originate from deep level defects or  nanostructuring (e.g., quantum dots). NaCl was not predicted as suitable for QT in the empirical approach.  

Out of the nine predicted materials (threshold of $0.85$) we note that both instances of CuI are stated in the MP database to decompose into trigonal CuI, which was not present in our predictions. Hexagonal CdSe decomposes into cubic CdSe (in the list of $9$), the cubic BN decomposes to hexagonal BN (mp-$984$, not in the predictions), and ZnCd$_3$Se$_4$ decomposes to ZnSe $+$ CdSe (in the list). Finally, both compounds of the BC$_2$N structure are listed to decompose into hexagonal BN $+$ C. 
However, synthesis may still be possible depending on the growth conditions. 

Lowering the cut-off requirement from $0.85$ to $0.75$ for all ML methods results in $66$ candidate materials for the empirical approach.  
The full list of these $66$ candidates is displayed in the Supplementary Information at \cite{supplementary}. 
In addition to the nine materials discussed above and some elemental and binary semiconductors, the list of $66$ predicted suitable candidates now also includes ternary compounds of the formula ABC$_2$. For the ABC$_2$ structures, the elements Ga, Cd, In and Zn can occupy the A-site, Cu, Sn, Ag and Ge take the B-site, while S, Se, Te, P or As may reside at the C site. Most of the predicted compounds include at least one toxic element with one exception: ZnGeP$_2$ (mp-$4524$) in a chalcopyrite-like tetragonal crystal structure with an indirect band gap of $1.2$~eV \cite{Zhang2015} as reported in the MP database. In comparison, the experimentally reported band gap is somewhat larger at $2.0$~eV \cite{Xing1989}. 
ZnGeP$_2$ crystallizes in a non-polar space group, possesses no magnetic moment, exhibits strongly covalent bonds, and has been reported as an excellent mid-IR transparent crystal material that is suitable for nonlinear optical applications \cite{Zhang2015}. Importantly, it is possible to integrate sources of photon quantum states based on nonlinear optics with ZnGeP$_2$ \cite{Caspani2017}. 
We therefore identify ZnGeP$_2$ as an eligible candidate material for QT, but it remains to be seen whether the candidate can facilitate, e.g., the isolated deep energy levels often associated with defects exhibiting quantum compatible properties, or instead be a candidate for nanostructure based QT.  

\begin{table}[t]
    \centering 
    \caption{Overview of material properties as listed in the MP database for the $9$ materials predicted by the empirical approach ($>0.85$) and the $6$ materials predicted by all approaches ($>0.75$). Compounds marked by an asterisk are predicted to be unstable to transformation into other phases in the MP database. }
    \begin{tabular}{c|c c c c c c c }
      & Material & Crystal structure & MP code & Density (g/cm$^3$) & Band gap (eV) & a, b, c (\AA) & $\alpha,\beta,\gamma$ ($^\circ$) \\
    \hline 
    Empirical & CdSe & Hexagonal & mp-$1070$ & $5.3$ & $0.6$ & $4.4,4.4,7.2$ & $90,90,120$ \\
    approach  & CuI & Hexagonal & mp-$569346$ & $5.8$ & $1.2$ & $4.3,4.3,7.0$ & $90,90,120$ \\ 
    to $0.85$  & CuI & Cubic & mp-$22895$ & $5.8$ & $1.2$ & $4.3,4.3,4.3$ & $60,60,60$  \\ 
    confidence & CdSe & Cubic & mp-$2691$ & $5.3$ & $0.5$ & $4.4,4.4,4.4$ & $60,60,60$  \\
     & BN & Cubic & mp-$1639$ & $3.5$ & $4.6$ & $2.6,2.6,2.6$ & $60,60,60$  \\
     & InAs & Cubic & mp-$20305$ & $5.3$ & $0.3$ & $4.4,4.4,4.4$ & $60,60,60$ \\
     & ZnCd$_3$Se$_4$ & Cubic & mp-$1078597$ & $5.3$ & $1.7$ & $6.1,6.1,6.1$ & $90,90,90$ \\
     & BC$_2$N & Tetragonal & mp-$1008523$ & $3.3$ & $1.6$ & $2.6,2.6,3.7$ & $90,90,90$ \\
     & BC$_2$N & Orthorombic & mp-$629458$ & $3.4$ & $1.8$ & $2.5,2.6,3.6$ & $90,90,90$ \\
    \hline 
    All & CdSnP$_2$ & Tetragonal & mp-$5213$ & $4.6$ & $0.7$ & $7.2,7.2,7.2$ & $131.1,131.1,71.7$ \\
    approaches & GeSe & Cubic & mp-$10759$ & $5.5$ & $0.4$ & $4.0,4.0,4.0$ & $60,60,60$ \\
    to $0.75$  & InP & Cubic & mp-$20351$ & $4.6$ & $0.5$ & $4.2,4.2,4.2$ &  $60,60,60$\\ 
    confidence & InP & Hexagonal & mp-$966800$ & $4.6$ & $0.5$ & $4.2,4.2,6.9$ & $90,90,120$ \\ 
     & SiSn & Cubic & mp-$1009813$ & $4.4$ & $0.4$ & $4.3,4.3,4.3$ & $60,60,60$ \\
     & ZnGeAs$_2$ & Tetragonal & mp-$4008$ & $5.2$ & $0.6$ & $7.0,7.0,7.0$ & $131.4,131.4,71.2$ \\
    \end{tabular}
    \label{tab:materialproperties}
\end{table}

Among the $66$ materials that all ML methods in the empirical approach agreed on to a $0.75$ confidence level, we also emphasize the presence of interesting materials like Ge (in three configurations mp-$137$, mp-$1067619$ and mp-$1198022$), GeC, BP and InP. The element Ge in the cubic structure (mp-$1198022$) shares many properties with Si and C in addition to the periodic column number. 
Germanium has the highest hole mobility of all semiconductors at room temperature, and is therefore considered a key material for the process of extending the chip performance in classical computers beyond the limits imposed by miniaturization \cite{Scappucci2020}. Furthermore, like SiC, device design based on Ge can take advantage of the mature large-scale fabrication of silicon due to the material's comparable properties.  
Similar considerations could be made for the case of GeC. 
Data from the MP database suggests that the cubic compound GeC (mp-$1002164$) is a covalently bonded semiconductor having a band gap of $1.8$~eV. 
Consequently, with SiC being widely known as a highly suitable host material for QT compatible defects, we encourage further research on GeC due to its similarities with SiC. 
Next, the compound BP (mp-$1479$ and mp-$1008559$) is present in our predictions in both the cubic and hexagonal structures, with indirect band gaps calculated at $1.5$~eV and $1.1$~eV, respectively. Both configurations of BP are nonmagnetic, non-toxic and BP has been synthesized with a potential for large-scale production \cite{MukhanovVladimirA2016Umso}. 
Lastly, InP (mp-$966800$) was also predicted as a suitable candidate to $0.75$ confidence. The compound inhabits the hexagonal structure with corner-sharing InP$_4$ tetrahedra. InP is reported in the Materials Project to have a direct band gap of $0.5$~eV and is considered as one of the most promising candidates to compete with Cd- or Pb- based QDs for, e.g., display and lighting applications  \cite{Zhang2020a, Won2019}. 
We underline the possibility of using InP-based quantum dots for QT applications. 

We note that all ML methods in the empirical approach agree on several oxides being potential candidates when the prediction cut-off is set low enough (but still above $0.5$). However, we find that almost all the oxide compounds fall between the decision boundary defining suitable and unsuitable candidates, and none are present in the list containing the $66$ suitable candidates using a $0.75$ threshold. Due to the labeling of ZnO as a suitable candidate in the training set, we believe that the boundary was shifted sufficiently to admit several oxides as suitable candidates. Removing ZnO from the training set for the empirical approach might result in fewer or even no oxide compounds being predicted above a suitability limit of $0.5$ and could be an interesting future pathway for investigation. 

Comparing to the work of \citeauthor{Ferrenti2020} \cite{Ferrenti2020}, they suggest a list of $541$ viable hosts after the data mining procedure.  
Among these, we find only a single material present in our list of $66$ candidates predicted by the four ML methods in the empirical approach: the nontoxic compound MgSe (mp-$10760$) which crystallizes in the rock-salt structure, is expected to have a $2.0$~eV band gap and decomposes to a similar MgSe configuration. 
MgSe is notable for its available spin-zero isotopes in accordance with the criteria set by \citeauthor{Weber2010} \cite{Weber2010} and \citeauthor{Ferrenti2020} \cite{Ferrenti2020}. We note that these properties may favor defects acting as spin centers with qubit potential and MgSe is thus identified as an interesting host material in this regard.   

\subsubsection*{Comparing the approaches}
Comparing the three approaches Ferrenti, extended Ferrenti and empirical (see also Table~\ref{tab:probabilites}), we observe that the extended Ferrenti approach is the least restricted also for the predicted materials and admits the largest number of entries regardless of the confidence threshold. 
The Ferrenti approach also predicts a large amount of materials as suitable compared to the empirical approach, but the resulting data are otherwise not very different from the extended version. The empirical approach, on the other hand, predicts fewer suitable candidates that are possible to examine  manually. 
Manual verification through a literature survey will often not be possible, however, and perfecting automatic data mining and analysis is therefore an important goal of material informatics \cite{rickman2019}. 

Despite notable differences, as mentioned above, there is substantial overlap of predicted materials between the three approaches. 
We find that $119$ of the $214$ candidates predicted as suitable by the empirical approach (threshold at $0.5$) were also selected by all ML methods in the extended Ferrenti approach. 
Similarly, $78$ of the $214$ materials are also predicted as suitable by all ML methods in the Ferrenti approach. All approaches and their corresponding ML methods agree on a total of $47$ potential candidates to $0.5$ confidence, where $8$ are elementary (unary), $29$ binary, and $10$ tertiary  
(see the Supplementary Information at \cite{supplementary} for the full list). 
Several interesting materials that were also discussed above for the predictions by the empirical approach are present among these $47$, including BP, CdSe, GeC, InP and Ge. However, we also find certain materials that cannot be classified as semiconductors, such as P, I, N$_2$ and H$_2$. We note that none of these were included by the empirical approach when the threshold was set to $0.75$ or above. 

Importantly, there are $6$ materials that all approaches (Ferrenti, extended Ferrenti and empirical) and ML methods agree on above a $0.75$ threshold. These are ZnGeAs$_2$ (tetragonal mp-$4008$), CdSnP$_2$ (tetragonal mp-$5213$), GeSe (cubic mp-$10759$), InP (cubic mp-$20351$), InP (hexagonal mp-$966800$) and SiSn (cubic mp-$1009813$). Here, we can distinguish three groupings in crystal structure and they agree with those discussed above; cubic, hexagonal and tetragonal (illustrated in Fig.~\ref{fig:crystalsystems}). The relevant material properties are summarized in the lower part of Table~\ref{tab:materialproperties}. 
Note that hexagonal InP is listed in the MP database as decomposing into cubic InP while SiSn is stated to decompose into Si $+$ Sn. 
We can deduce a series of similar trends as for the $9$ materials from the empirical approach --- e.g., the same space groups are manifesting, we find only binary or tertiary compounds, and $a=b$ for all materials while $c$ is different for hexagonal InP (unstable). However, consistent four-fold coordination and the bond angle trends we observed for the first $9$ materials are not maintained as GeSe is six-fold coordinated. 
We highlight these six compounds, along with the nine predicted by the empirical approach to $0.85$ confidence, as particularly interesting for future in-depth theoretical and experimental studies. 

\section*{Discussion}

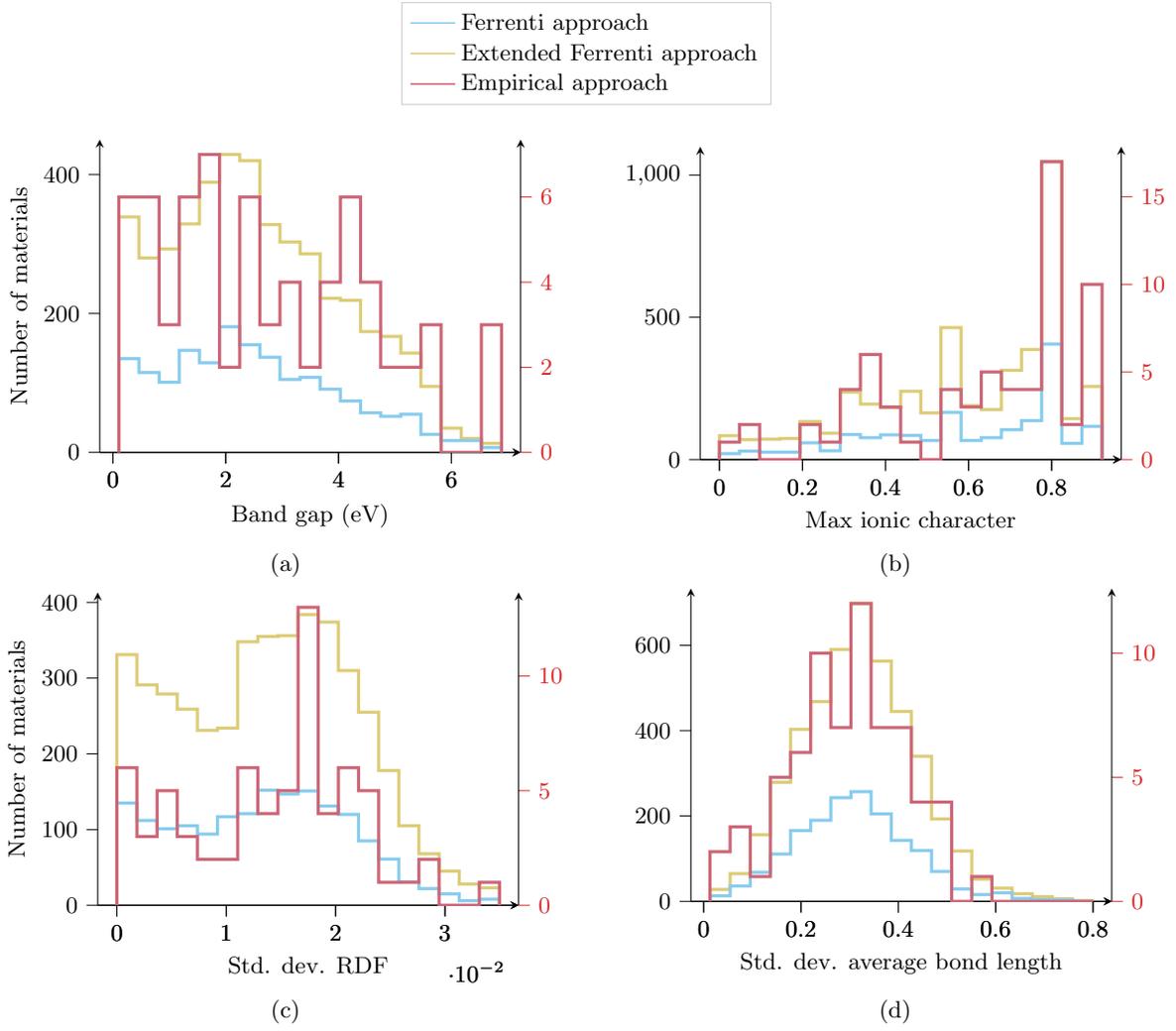
\begin{figure}[ht]
    \begin{subfigure}[b]{1\textwidth}
    \centering
\begin{tikzpicture}

\definecolor{color0}{rgb}{0.866666666666667,0.8,0.466666666666667}
\definecolor{color1}{rgb}{0.533333333333333,0.8,0.933333333333333}
\definecolor{color3}{rgb}{0.8,0.4,0.466666666666667}
\begin{axis}[%
 hide axis,
 xmin=0,
 xmax=1,
 ymin=0,
 ymax=1,
 legend columns=1,
 legend cell align={left},
 legend style={
   fill opacity=1,
   draw opacity=1,
   text opacity=1,
   align=center,
   anchor=north,
   draw=white!80!black
 },
 ]
 \addlegendimage{semithick, color1}
 \addlegendentry{Ferrenti approach};
 \addlegendimage{semithick, color0}
 \addlegendentry{Extended Ferrenti approach};
 \addlegendimage{semithick, color3}
 \addlegendentry{Empirical approach};
 \end{axis}

\end{tikzpicture}
  \end{subfigure}
  \par\bigskip
\begin{subfigure}[b]{0.45\textwidth}
\begin{tikzpicture}

\definecolor{color0}{rgb}{0.866666666666667,0.8,0.466666666666667}
\definecolor{color1}{rgb}{0.533333333333333,0.8,0.933333333333333}
\definecolor{color2}{rgb}{0.83921568627451,0.152941176470588,0.156862745098039}
\definecolor{color3}{rgb}{0.8,0.4,0.466666666666667}

\begin{axis}[
height=2.275590092707901in,
legend cell align={left},
legend style={fill opacity=0.8, draw opacity=1, text opacity=1, draw=white!80!black, font=\footnotesize},
tick align=outside,
tick pos=left,
title={},
width=2.8444876158848764in,
x grid style={white!69.0196078431373!black},
xlabel={Band gap (eV)},
xmin=-0.234255, xmax=7.229355,
xtick style={color=black},
y grid style={white!69.0196078431373!black},
ylabel={Number of materials},
ymin=0, ymax=450.45,
ytick style={color=black}
]
\path [draw=color0, very thick]
(axis cs:0.105,0)
--(axis cs:0.105,339)
--(axis cs:0.462110526315789,339)
--(axis cs:0.462110526315789,280)
--(axis cs:0.819221052631579,280)
--(axis cs:0.819221052631579,293)
--(axis cs:1.17633157894737,293)
--(axis cs:1.17633157894737,329)
--(axis cs:1.53344210526316,329)
--(axis cs:1.53344210526316,389)
--(axis cs:1.89055263157895,389)
--(axis cs:1.89055263157895,429)
--(axis cs:2.24766315789474,429)
--(axis cs:2.24766315789474,420)
--(axis cs:2.60477368421053,420)
--(axis cs:2.60477368421053,328)
--(axis cs:2.96188421052632,328)
--(axis cs:2.96188421052632,303)
--(axis cs:3.31899473684211,303)
--(axis cs:3.31899473684211,286)
--(axis cs:3.67610526315789,286)
--(axis cs:3.67610526315789,222)
--(axis cs:4.03321578947368,222)
--(axis cs:4.03321578947368,219)
--(axis cs:4.39032631578947,219)
--(axis cs:4.39032631578947,174)
--(axis cs:4.74743684210526,174)
--(axis cs:4.74743684210526,167)
--(axis cs:5.10454736842105,167)
--(axis cs:5.10454736842105,143)
--(axis cs:5.46165789473684,143)
--(axis cs:5.46165789473684,95)
--(axis cs:5.81876842105263,95)
--(axis cs:5.81876842105263,35)
--(axis cs:6.17587894736842,35)
--(axis cs:6.17587894736842,20)
--(axis cs:6.53298947368421,20)
--(axis cs:6.53298947368421,13)
--(axis cs:6.8901,13)
--(axis cs:6.8901,0);

\path [draw=color1, very thick]
(axis cs:0.105,0)
--(axis cs:0.105,135)
--(axis cs:0.462110526315789,135)
--(axis cs:0.462110526315789,115)
--(axis cs:0.819221052631579,115)
--(axis cs:0.819221052631579,101)
--(axis cs:1.17633157894737,101)
--(axis cs:1.17633157894737,147)
--(axis cs:1.53344210526316,147)
--(axis cs:1.53344210526316,129)
--(axis cs:1.89055263157895,129)
--(axis cs:1.89055263157895,181)
--(axis cs:2.24766315789474,181)
--(axis cs:2.24766315789474,155)
--(axis cs:2.60477368421053,155)
--(axis cs:2.60477368421053,137)
--(axis cs:2.96188421052632,137)
--(axis cs:2.96188421052632,105)
--(axis cs:3.31899473684211,105)
--(axis cs:3.31899473684211,108)
--(axis cs:3.67610526315789,108)
--(axis cs:3.67610526315789,91)
--(axis cs:4.03321578947368,91)
--(axis cs:4.03321578947368,74)
--(axis cs:4.39032631578947,74)
--(axis cs:4.39032631578947,57)
--(axis cs:4.74743684210526,57)
--(axis cs:4.74743684210526,52)
--(axis cs:5.10454736842105,52)
--(axis cs:5.10454736842105,55)
--(axis cs:5.46165789473684,55)
--(axis cs:5.46165789473684,26)
--(axis cs:5.81876842105263,26)
--(axis cs:5.81876842105263,17)
--(axis cs:6.17587894736842,17)
--(axis cs:6.17587894736842,17)
--(axis cs:6.53298947368421,17)
--(axis cs:6.53298947368421,7)
--(axis cs:6.8901,7)
--(axis cs:6.8901,0);

\end{axis}

\begin{axis}[
axis y line=right,
height=2.275590092707901in,
tick align=outside,
width=2.8444876158848764in,
x grid style={white!69.0196078431373!black},
xmin=-0.234255, xmax=7.229355,
xtick pos=left,
xtick style={color=black},
y grid style={white!69.0196078431373!black},
ymin=0, ymax=7.35,
ytick pos=right,
ytick style={color=color2},
yticklabel style={anchor=west, color=color2}
]
\path [draw=color3, very thick]
(axis cs:0.105,0)
--(axis cs:0.105,6)
--(axis cs:0.462110526315789,6)
--(axis cs:0.462110526315789,6)
--(axis cs:0.819221052631579,6)
--(axis cs:0.819221052631579,3)
--(axis cs:1.17633157894737,3)
--(axis cs:1.17633157894737,6)
--(axis cs:1.53344210526316,6)
--(axis cs:1.53344210526316,7)
--(axis cs:1.89055263157895,7)
--(axis cs:1.89055263157895,2)
--(axis cs:2.24766315789474,2)
--(axis cs:2.24766315789474,6)
--(axis cs:2.60477368421053,6)
--(axis cs:2.60477368421053,3)
--(axis cs:2.96188421052632,3)
--(axis cs:2.96188421052632,4)
--(axis cs:3.31899473684211,4)
--(axis cs:3.31899473684211,2)
--(axis cs:3.67610526315789,2)
--(axis cs:3.67610526315789,4)
--(axis cs:4.03321578947368,4)
--(axis cs:4.03321578947368,6)
--(axis cs:4.39032631578947,6)
--(axis cs:4.39032631578947,4)
--(axis cs:4.74743684210526,4)
--(axis cs:4.74743684210526,2)
--(axis cs:5.10454736842105,2)
--(axis cs:5.10454736842105,2)
--(axis cs:5.46165789473684,2)
--(axis cs:5.46165789473684,3)
--(axis cs:5.81876842105263,3)
--(axis cs:5.81876842105263,0)
--(axis cs:6.17587894736842,0)
--(axis cs:6.17587894736842,0)
--(axis cs:6.53298947368421,0)
--(axis cs:6.53298947368421,3)
--(axis cs:6.8901,3)
--(axis cs:6.8901,0);
\end{axis}

\end{tikzpicture}
    \subcaption{}
\end{subfigure}
\begin{subfigure}[b]{0.45\textwidth}
\begin{tikzpicture}

\definecolor{color0}{rgb}{0.866666666666667,0.8,0.466666666666667}
\definecolor{color1}{rgb}{0.533333333333333,0.8,0.933333333333333}
\definecolor{color2}{rgb}{0.83921568627451,0.152941176470588,0.156862745098039}
\definecolor{color3}{rgb}{0.8,0.4,0.466666666666667}

\begin{axis}[
height=2.275590092707901in,
legend cell align={left},
legend style={fill opacity=0.8, draw opacity=1, text opacity=1, draw=white!80!black},
tick align=outside,
tick pos=left,
title={},
width=2.8444876158848764in,
x grid style={white!69.0196078431373!black},
xlabel={Max ionic character},
xmin=-0.0460725199741125, xmax=0.967522919456363,
xtick style={color=black},
y grid style={white!69.0196078431373!black},
ylabel={},
ymin=0, ymax=1099.35,
ytick style={color=black}
]
\path [draw=color0, very thick]
(axis cs:0,0)
--(axis cs:0,84)
--(axis cs:0.0484973894464342,84)
--(axis cs:0.0484973894464342,70)
--(axis cs:0.0969947788928685,70)
--(axis cs:0.0969947788928685,72)
--(axis cs:0.145492168339303,72)
--(axis cs:0.145492168339303,74)
--(axis cs:0.193989557785737,74)
--(axis cs:0.193989557785737,134)
--(axis cs:0.242486947232171,134)
--(axis cs:0.242486947232171,93)
--(axis cs:0.290984336678605,93)
--(axis cs:0.290984336678605,237)
--(axis cs:0.33948172612504,237)
--(axis cs:0.33948172612504,195)
--(axis cs:0.387979115571474,195)
--(axis cs:0.387979115571474,183)
--(axis cs:0.436476505017908,183)
--(axis cs:0.436476505017908,240)
--(axis cs:0.484973894464342,240)
--(axis cs:0.484973894464342,164)
--(axis cs:0.533471283910777,164)
--(axis cs:0.533471283910777,464)
--(axis cs:0.581968673357211,464)
--(axis cs:0.581968673357211,190)
--(axis cs:0.630466062803645,190)
--(axis cs:0.630466062803645,176)
--(axis cs:0.678963452250079,176)
--(axis cs:0.678963452250079,314)
--(axis cs:0.727460841696513,314)
--(axis cs:0.727460841696513,387)
--(axis cs:0.775958231142948,387)
--(axis cs:0.775958231142948,1047)
--(axis cs:0.824455620589382,1047)
--(axis cs:0.824455620589382,144)
--(axis cs:0.872953010035816,144)
--(axis cs:0.872953010035816,257)
--(axis cs:0.92145039948225,257)
--(axis cs:0.92145039948225,0);

\path [draw=color1, very thick]
(axis cs:0,0)
--(axis cs:0,21)
--(axis cs:0.0484973894464342,21)
--(axis cs:0.0484973894464342,30)
--(axis cs:0.0969947788928685,30)
--(axis cs:0.0969947788928685,26)
--(axis cs:0.145492168339303,26)
--(axis cs:0.145492168339303,26)
--(axis cs:0.193989557785737,26)
--(axis cs:0.193989557785737,59)
--(axis cs:0.242486947232171,59)
--(axis cs:0.242486947232171,31)
--(axis cs:0.290984336678605,31)
--(axis cs:0.290984336678605,88)
--(axis cs:0.33948172612504,88)
--(axis cs:0.33948172612504,77)
--(axis cs:0.387979115571474,77)
--(axis cs:0.387979115571474,87)
--(axis cs:0.436476505017908,87)
--(axis cs:0.436476505017908,85)
--(axis cs:0.484973894464342,85)
--(axis cs:0.484973894464342,67)
--(axis cs:0.533471283910777,67)
--(axis cs:0.533471283910777,166)
--(axis cs:0.581968673357211,166)
--(axis cs:0.581968673357211,67)
--(axis cs:0.630466062803645,67)
--(axis cs:0.630466062803645,77)
--(axis cs:0.678963452250079,77)
--(axis cs:0.678963452250079,105)
--(axis cs:0.727460841696513,105)
--(axis cs:0.727460841696513,137)
--(axis cs:0.775958231142948,137)
--(axis cs:0.775958231142948,406)
--(axis cs:0.824455620589382,406)
--(axis cs:0.824455620589382,57)
--(axis cs:0.872953010035816,57)
--(axis cs:0.872953010035816,117)
--(axis cs:0.92145039948225,117)
--(axis cs:0.92145039948225,0);

\end{axis}

\begin{axis}[
axis y line=right,
height=2.275590092707901in,
tick align=outside,
width=2.8444876158848764in,
x grid style={white!69.0196078431373!black},
xmin=-0.0460725199741125, xmax=0.967522919456363,
xtick pos=left,
xtick style={color=black},
y grid style={white!69.0196078431373!black},
ymin=0, ymax=17.85,
ytick pos=right,
ytick style={color=color2},
yticklabel style={anchor=west, color=color2}
]
\path [draw=color3, very thick]
(axis cs:0,0)
--(axis cs:0,1)
--(axis cs:0.0484973894464342,1)
--(axis cs:0.0484973894464342,2)
--(axis cs:0.0969947788928685,2)
--(axis cs:0.0969947788928685,0)
--(axis cs:0.145492168339303,0)
--(axis cs:0.145492168339303,0)
--(axis cs:0.193989557785737,0)
--(axis cs:0.193989557785737,2)
--(axis cs:0.242486947232171,2)
--(axis cs:0.242486947232171,1)
--(axis cs:0.290984336678605,1)
--(axis cs:0.290984336678605,4)
--(axis cs:0.33948172612504,4)
--(axis cs:0.33948172612504,6)
--(axis cs:0.387979115571474,6)
--(axis cs:0.387979115571474,3)
--(axis cs:0.436476505017908,3)
--(axis cs:0.436476505017908,1)
--(axis cs:0.484973894464342,1)
--(axis cs:0.484973894464342,0)
--(axis cs:0.533471283910777,0)
--(axis cs:0.533471283910777,4)
--(axis cs:0.581968673357211,4)
--(axis cs:0.581968673357211,3)
--(axis cs:0.630466062803645,3)
--(axis cs:0.630466062803645,5)
--(axis cs:0.678963452250079,5)
--(axis cs:0.678963452250079,4)
--(axis cs:0.727460841696513,4)
--(axis cs:0.727460841696513,4)
--(axis cs:0.775958231142948,4)
--(axis cs:0.775958231142948,17)
--(axis cs:0.824455620589382,17)
--(axis cs:0.824455620589382,2)
--(axis cs:0.872953010035816,2)
--(axis cs:0.872953010035816,10)
--(axis cs:0.92145039948225,10)
--(axis cs:0.92145039948225,0);
\end{axis}

\end{tikzpicture}

    \subcaption{}
\end{subfigure}%

\begin{subfigure}[b]{0.45\textwidth}
\begin{tikzpicture}

\definecolor{color0}{rgb}{0.866666666666667,0.8,0.466666666666667}
\definecolor{color1}{rgb}{0.533333333333333,0.8,0.933333333333333}
\definecolor{color2}{rgb}{0.83921568627451,0.152941176470588,0.156862745098039}
\definecolor{color3}{rgb}{0.8,0.4,0.466666666666667}

\begin{axis}[
height=2.275590092707901in,
legend cell align={left},
legend style={fill opacity=0.8, draw opacity=1, text opacity=1, draw=white!80!black},
tick align=outside,
tick pos=left,
title={},
width=2.8444876158848764in,
x grid style={white!69.0196078431373!black},
xlabel={Std. dev. RDF},
xmin=-0.00174896633573773, xmax=0.0367282930504924,
xtick style={color=black},
y grid style={white!69.0196078431373!black},
ylabel={Number of materials},
ymin=0, ymax=413.2,
ytick style={color=black}
]
\path [draw=color0, very thick]
(axis cs:0,0)
--(axis cs:0,331)
--(axis cs:0.0018410171955134,331)
--(axis cs:0.0018410171955134,291)
--(axis cs:0.00368203439102681,291)
--(axis cs:0.00368203439102681,279)
--(axis cs:0.00552305158654021,279)
--(axis cs:0.00552305158654021,259)
--(axis cs:0.00736406878205362,259)
--(axis cs:0.00736406878205362,231)
--(axis cs:0.00920508597756702,231)
--(axis cs:0.00920508597756702,234)
--(axis cs:0.0110461031730804,234)
--(axis cs:0.0110461031730804,348)
--(axis cs:0.0128871203685938,348)
--(axis cs:0.0128871203685938,355)
--(axis cs:0.0147281375641072,355)
--(axis cs:0.0147281375641072,356)
--(axis cs:0.0165691547596206,356)
--(axis cs:0.0165691547596206,384)
--(axis cs:0.018410171955134,384)
--(axis cs:0.018410171955134,374)
--(axis cs:0.0202511891506474,374)
--(axis cs:0.0202511891506474,310)
--(axis cs:0.0220922063461609,310)
--(axis cs:0.0220922063461609,255)
--(axis cs:0.0239332235416743,255)
--(axis cs:0.0239332235416743,178)
--(axis cs:0.0257742407371877,178)
--(axis cs:0.0257742407371877,105)
--(axis cs:0.0276152579327011,105)
--(axis cs:0.0276152579327011,68)
--(axis cs:0.0294562751282145,68)
--(axis cs:0.0294562751282145,45)
--(axis cs:0.0312972923237279,45)
--(axis cs:0.0312972923237279,28)
--(axis cs:0.0331383095192413,28)
--(axis cs:0.0331383095192413,23)
--(axis cs:0.0349793267147547,23)
--(axis cs:0.0349793267147547,0);

\path [draw=color1, very thick]
(axis cs:0,0)
--(axis cs:0,135)
--(axis cs:0.0018410171955134,135)
--(axis cs:0.0018410171955134,112)
--(axis cs:0.00368203439102681,112)
--(axis cs:0.00368203439102681,101)
--(axis cs:0.00552305158654021,101)
--(axis cs:0.00552305158654021,105)
--(axis cs:0.00736406878205362,105)
--(axis cs:0.00736406878205362,94)
--(axis cs:0.00920508597756702,94)
--(axis cs:0.00920508597756702,117)
--(axis cs:0.0110461031730804,117)
--(axis cs:0.0110461031730804,121)
--(axis cs:0.0128871203685938,121)
--(axis cs:0.0128871203685938,152)
--(axis cs:0.0147281375641072,152)
--(axis cs:0.0147281375641072,147)
--(axis cs:0.0165691547596206,147)
--(axis cs:0.0165691547596206,151)
--(axis cs:0.018410171955134,151)
--(axis cs:0.018410171955134,131)
--(axis cs:0.0202511891506474,131)
--(axis cs:0.0202511891506474,120)
--(axis cs:0.0220922063461609,120)
--(axis cs:0.0220922063461609,85)
--(axis cs:0.0239332235416743,85)
--(axis cs:0.0239332235416743,61)
--(axis cs:0.0257742407371877,61)
--(axis cs:0.0257742407371877,31)
--(axis cs:0.0276152579327011,31)
--(axis cs:0.0276152579327011,22)
--(axis cs:0.0294562751282145,22)
--(axis cs:0.0294562751282145,15)
--(axis cs:0.0312972923237279,15)
--(axis cs:0.0312972923237279,6)
--(axis cs:0.0331383095192413,6)
--(axis cs:0.0331383095192413,8)
--(axis cs:0.0349793267147547,8)
--(axis cs:0.0349793267147547,0);
\end{axis}

\begin{axis}[
axis y line=right,
height=2.275590092707901in,
tick align=outside,
width=2.8444876158848764in,
x grid style={white!69.0196078431373!black},
xmin=-0.00174896633573773, xmax=0.0367282930504924,
xtick pos=left,
xtick style={color=black},
y grid style={white!69.0196078431373!black},
ymin=0, ymax=13.65,
ytick pos=right,
ytick style={color=color2},
yticklabel style={anchor=west, color=color2}
]
\path [draw=color3, very thick]
(axis cs:0,0)
--(axis cs:0,6)
--(axis cs:0.0018410171955134,6)
--(axis cs:0.0018410171955134,3)
--(axis cs:0.00368203439102681,3)
--(axis cs:0.00368203439102681,5)
--(axis cs:0.00552305158654021,5)
--(axis cs:0.00552305158654021,3)
--(axis cs:0.00736406878205362,3)
--(axis cs:0.00736406878205362,2)
--(axis cs:0.00920508597756702,2)
--(axis cs:0.00920508597756702,2)
--(axis cs:0.0110461031730804,2)
--(axis cs:0.0110461031730804,6)
--(axis cs:0.0128871203685938,6)
--(axis cs:0.0128871203685938,4)
--(axis cs:0.0147281375641072,4)
--(axis cs:0.0147281375641072,5)
--(axis cs:0.0165691547596206,5)
--(axis cs:0.0165691547596206,13)
--(axis cs:0.018410171955134,13)
--(axis cs:0.018410171955134,4)
--(axis cs:0.0202511891506474,4)
--(axis cs:0.0202511891506474,6)
--(axis cs:0.0220922063461609,6)
--(axis cs:0.0220922063461609,5)
--(axis cs:0.0239332235416743,5)
--(axis cs:0.0239332235416743,1)
--(axis cs:0.0257742407371877,1)
--(axis cs:0.0257742407371877,1)
--(axis cs:0.0276152579327011,1)
--(axis cs:0.0276152579327011,2)
--(axis cs:0.0294562751282145,2)
--(axis cs:0.0294562751282145,0)
--(axis cs:0.0312972923237279,0)
--(axis cs:0.0312972923237279,0)
--(axis cs:0.0331383095192413,0)
--(axis cs:0.0331383095192413,1)
--(axis cs:0.0349793267147547,1)
--(axis cs:0.0349793267147547,0);
\end{axis}

\end{tikzpicture}

    \subcaption{}
\end{subfigure}
\begin{subfigure}[b]{0.45\textwidth}
\begin{tikzpicture}

\definecolor{color0}{rgb}{0.866666666666667,0.8,0.466666666666667}
\definecolor{color1}{rgb}{0.533333333333333,0.8,0.933333333333333}
\definecolor{color2}{rgb}{0.83921568627451,0.152941176470588,0.156862745098039}
\definecolor{color3}{rgb}{0.8,0.4,0.466666666666667}

\begin{axis}[
height=2.275590092707901in,
legend cell align={left},
legend style={fill opacity=0.8, draw opacity=1, text opacity=1, draw=white!80!black},
tick align=outside,
tick pos=left,
title={},
width=2.8444876158848764in,
x grid style={white!69.0196078431373!black},
xlabel={Std. dev. average bond length},
xmin=-0.0263144046185111, xmax=0.838491250528538,
xtick style={color=black},
y grid style={white!69.0196078431373!black},
ylabel={},
ymin=0, ymax=732.9,
ytick style={color=black}
]
\path [draw=color0, very thick]
(axis cs:0.0129949433427184,0)
--(axis cs:0.0129949433427184,28)
--(axis cs:0.0543732043545389,28)
--(axis cs:0.0543732043545389,65)
--(axis cs:0.0957514653663595,65)
--(axis cs:0.0957514653663595,156)
--(axis cs:0.13712972637818,156)
--(axis cs:0.13712972637818,279)
--(axis cs:0.178507987390001,279)
--(axis cs:0.178507987390001,403)
--(axis cs:0.219886248401821,403)
--(axis cs:0.219886248401821,468)
--(axis cs:0.261264509413642,468)
--(axis cs:0.261264509413642,590)
--(axis cs:0.302642770425462,590)
--(axis cs:0.302642770425462,698)
--(axis cs:0.344021031437283,698)
--(axis cs:0.344021031437283,563)
--(axis cs:0.385399292449103,563)
--(axis cs:0.385399292449103,445)
--(axis cs:0.426777553460924,445)
--(axis cs:0.426777553460924,340)
--(axis cs:0.468155814472744,340)
--(axis cs:0.468155814472744,193)
--(axis cs:0.509534075484565,193)
--(axis cs:0.509534075484565,118)
--(axis cs:0.550912336496385,118)
--(axis cs:0.550912336496385,52)
--(axis cs:0.592290597508206,52)
--(axis cs:0.592290597508206,31)
--(axis cs:0.633668858520026,31)
--(axis cs:0.633668858520026,18)
--(axis cs:0.675047119531847,18)
--(axis cs:0.675047119531847,11)
--(axis cs:0.716425380543667,11)
--(axis cs:0.716425380543667,6)
--(axis cs:0.757803641555488,6)
--(axis cs:0.757803641555488,2)
--(axis cs:0.799181902567308,2)
--(axis cs:0.799181902567308,0);

\path [draw=color1, very thick]
(axis cs:0.0129949433427184,0)
--(axis cs:0.0129949433427184,13)
--(axis cs:0.0543732043545389,13)
--(axis cs:0.0543732043545389,36)
--(axis cs:0.0957514653663595,36)
--(axis cs:0.0957514653663595,68)
--(axis cs:0.13712972637818,68)
--(axis cs:0.13712972637818,111)
--(axis cs:0.178507987390001,111)
--(axis cs:0.178507987390001,166)
--(axis cs:0.219886248401821,166)
--(axis cs:0.219886248401821,190)
--(axis cs:0.261264509413642,190)
--(axis cs:0.261264509413642,243)
--(axis cs:0.302642770425462,243)
--(axis cs:0.302642770425462,257)
--(axis cs:0.344021031437283,257)
--(axis cs:0.344021031437283,205)
--(axis cs:0.385399292449103,205)
--(axis cs:0.385399292449103,143)
--(axis cs:0.426777553460924,143)
--(axis cs:0.426777553460924,119)
--(axis cs:0.468155814472744,119)
--(axis cs:0.468155814472744,70)
--(axis cs:0.509534075484565,70)
--(axis cs:0.509534075484565,29)
--(axis cs:0.550912336496385,29)
--(axis cs:0.550912336496385,16)
--(axis cs:0.592290597508206,16)
--(axis cs:0.592290597508206,20)
--(axis cs:0.633668858520026,20)
--(axis cs:0.633668858520026,7)
--(axis cs:0.675047119531847,7)
--(axis cs:0.675047119531847,4)
--(axis cs:0.716425380543667,4)
--(axis cs:0.716425380543667,3)
--(axis cs:0.757803641555488,3)
--(axis cs:0.757803641555488,0)
--(axis cs:0.799181902567308,0)
--(axis cs:0.799181902567308,0);

\end{axis}

\begin{axis}[
axis y line=right,
height=2.275590092707901in,
tick align=outside,
width=2.8444876158848764in,
x grid style={white!69.0196078431373!black},
xmin=-0.0263144046185111, xmax=0.838491250528538,
xtick pos=left,
xtick style={color=black},
y grid style={white!69.0196078431373!black},
ymin=0, ymax=12.6,
ytick pos=right,
ytick style={color=color2},
yticklabel style={anchor=west, color=color2}
]
\path [draw=color3, very thick]
(axis cs:0.0129949433427184,0)
--(axis cs:0.0129949433427184,2)
--(axis cs:0.0543732043545389,2)
--(axis cs:0.0543732043545389,3)
--(axis cs:0.0957514653663595,3)
--(axis cs:0.0957514653663595,1)
--(axis cs:0.13712972637818,1)
--(axis cs:0.13712972637818,5)
--(axis cs:0.178507987390001,5)
--(axis cs:0.178507987390001,6)
--(axis cs:0.219886248401821,6)
--(axis cs:0.219886248401821,10)
--(axis cs:0.261264509413642,10)
--(axis cs:0.261264509413642,7)
--(axis cs:0.302642770425462,7)
--(axis cs:0.302642770425462,12)
--(axis cs:0.344021031437283,12)
--(axis cs:0.344021031437283,7)
--(axis cs:0.385399292449103,7)
--(axis cs:0.385399292449103,7)
--(axis cs:0.426777553460924,7)
--(axis cs:0.426777553460924,4)
--(axis cs:0.468155814472744,4)
--(axis cs:0.468155814472744,4)
--(axis cs:0.509534075484565,4)
--(axis cs:0.509534075484565,0)
--(axis cs:0.550912336496385,0)
--(axis cs:0.550912336496385,1)
--(axis cs:0.592290597508206,1)
--(axis cs:0.592290597508206,0)
--(axis cs:0.633668858520026,0)
--(axis cs:0.633668858520026,0)
--(axis cs:0.675047119531847,0)
--(axis cs:0.675047119531847,0)
--(axis cs:0.716425380543667,0)
--(axis cs:0.716425380543667,0)
--(axis cs:0.757803641555488,0)
--(axis cs:0.757803641555488,0)
--(axis cs:0.799181902567308,0)
--(axis cs:0.799181902567308,0);
\end{axis}

\end{tikzpicture}

    \subcaption{}
\end{subfigure}
\caption{Histograms of predicted suitable materials as a function of the (a) PBE-calculated band gap from Materials Project, (b) maximum ionic character, (c) standard deviation of the radial distribution function (RDF) and (d) standard deviation of the average bond length. The total number of predicted materials is  $6804$ for the Ferrenti approach, $9227$ for the extended Ferrenti approach, and $214$ for the empirical approach. The Ferrenti and extended Ferrenti approaches refer to the left $y$-axis and the empirical approach to the right.
    }
\label{fig:histogram_new}
\end{figure}

Taking a closer look at the reasoning behind the choices made by the different ML methods during the classification process, we can start to identify important driving forces for manifestation of quantum compatible properties in semiconductors. 
The analysis of the principal components extracted from the ML methods revealed that the most important principal component for both the Ferrenti and extended Ferrenti approaches encompasses features related to the band gap and chemical environment. This means that the band gap criterion imposed in the training set selection is at least somewhat satisfied. In other words, the Ferrenti approaches appear to reproduce the logic of the initial selection process. For the empirical approach, on the other hand, band gap related features were not recognized as important in the dominant principal component. 

Figure~\ref{fig:histogram_new}(a) displays the number of predicted suitable materials as a function of band gap (from the MP database), and reveals that all approaches predicted a substantial amount of materials with a low band gap (below $0.5$~eV). Moreover, the distribution of materials that were identified as suitable is rather broad with the Ferrenti approaches exhibiting a peak around a $2.5$~eV band gap. 
Coincidentally, the band gap of, e.g., 4$H$-SiC is usually computed at around $2.5$~eV using the PBE functional. 
The empirical approach exhibits a more scattered data distribution with local maxima around $1$-$2$ and $4$~eV. 

The secondmost important principal component in the Ferrenti approach encompasses properties such as the ionic character, covalent radius and maximum packing efficiency (see the Supplementary Information at \cite{supplementary} for the predicted material distributions of the latter two features).  
Intriguingly, as shown in Figure~\ref{fig:histogram_new}(b), the predicted candidates distribute over a broad range of ionic characters for all three approaches - even peaking at a relatively high ionic character of $0.8$. 
We note that this may be a result of the distribution of the initial dataset of $25.000$ materials that peaks around a similar value (not shown). The minor peak in the empirical approach's predicted materials around $0.4$ ionic character is not present for the Ferrenti approach, extended Ferrenti approach nor the overall data distribution. For comparison, all SiC entries in the dataset have maximum ionic characters of $\sim 0.1$. 
Further, we find that the covalent radii of the materials (see the Supplementary Information) exhibit two distinct peaks in the data distribution. 
The trend of two data peaks is repeated for the maximum packing efficiency but is much more prominent for the empirical approach. This indicates that the material density, or in other words the bond length, is an important parameter for QT suitability.  

The ML methods in the empirical approach consistently identified the first principal component as the predominant one. Identifying the single most important feature in this principal component proved challenging as it is the combined impact of several features that is important. 
Here, we find that the standard deviation of the radial distribution function (RDF) has a  particularly strong impact since it appears four times in different forms in the top ten list over dominating features. One configuration of the standard deviation of the RDF is demonstrated in Fig.~\ref{fig:histogram_new}(c), with two others being included in the Supplementary Information at \cite{supplementary}. Intriguingly, the standard deviation of the RDF exhibits substantial discrepancies between the Ferrenti approaches and the empirical approach. This reflects the observations from the principal component analysis where the first principal component is dominated by the RDF and similarly symmetry related features. For the empirical approach, there is a sharp peak in the preferred value for the standard deviation for the RDF, while the Ferrenti and extended Ferrenti approaches display an even distribution across a broader range. 

We interpret the standard deviation  of the RDF such that zero standard deviation in the RDF means that there is no variation in the radial symmetry throughout the material. Similarly, zero standard deviation in the average bond length would mean that all bonds are identical throughout the crystal. Intriguingly, the peaks in the distributions of the predicted materials are not found for perfectly symmetric materials with identical bond lengths; instead, some variation in the bond and wavefunction distributions is found to be optimal for a material to be suitable for QT.  
Note that the maxima in Fig.~\ref{fig:histogram_new}(c) seem to appear for moderate standard deviations in the RDF, indicating that a certain degree of symmetry is necessary for a material to act as a suitable QT host. The exact degree and type of symmetry is still open to debate and merits further study. 
Similar symmetry related features such as the standard deviation of the average bond length (see Fig.~\ref{fig:histogram_new}d), the site fingerprint of the chemical environment and the bond orientation (see the Supplementary Information at \cite{supplementary}) are also influential in guiding the ML algorithms upon classifying suitable and unsuitable materials for QT in the empirical approach.

Interestingly, none of the nine materials predicted by the empirical approach above a $0.85$ threshold, nor of the six materials predicted by all approaches to $0.75$ confidence, are elemental. At most three different elements are present, but the emphasis is clearly on binary compounds. This resonates with the observations from the principal component analysis; an optimal degree of crystalline order likely exists for a material to manifest QT compatible properties, but some variations throughout the crystal in, e.g., bond length and symmetry are needed. This is in contrast to the expectations that guided the formulation of the test and training sets in the Ferrenti and extended Ferrenti approaches. Where most previous works have highlighted features such as band gap, polarity and ionic character as vital for a semiconductor to manifest quantum compatible features, our results reveal that local variations in the crystal structure related to symmetry and bond angles, length and orientation should be considered as well and may be more important. 

Considering the trends in machine learning selectivity in light of the specific defect centers we know to be quantum compatible reveals fascinating characteristics. None of the known quantum emitters or spin centers seem to appear in completely uniform systems. For example, in the high symmetry materials of diamond and silicon, QT compatible defects are not intrinsic; neither the silicon vacancy in silicon nor the carbon vacancy in diamond, for example,  have exhibited single-photon emission or controllable spin coherence. Instead, we find that quantum effects often appear after impurities are introduced, as evidenced by the phosphorous and carbon impurities in silicon \cite{He2019,Redjem2020} and the nitrogen-vacancy, germanium-vacancy, tin-vacancy and lead-vacancy centers in diamond \cite{Thiering2020}. 
For the binary system of silicon carbide in different polytypes, on the other hand, intrinsic defects like the silicon vacancy and the divacancy are appropriate for our goals. 
While these trends have yet to be verified on a grander scale, our findings provide strong indications that local variations in the crystal structure are paramount for QT compatible properties to manifest in a semiconductor host. 

Our findings are corroborated by the fact that all four ML methods predict a small set of the unlabeled materials as suitable, while agreeing on a large part of these materials. 
The methods we have chosen are all well tested, with random forests and gradient boosting methods tending to outperform the others, resulting normally in a small bias and variance \cite{Hastie2009,Mehta2019,Murphy2012}. 

To summarize, the data extraction tools developed in this work resulted in a dataset of over $25.000$ materials and more than $4800$ physics-informed features. Three different approaches (Ferrenti, extended Ferrenti and empirical) for data labeling were employed for training and testing of four ML algorithms. 
We find a much clearer separation between suitable and unsuitable candidates after data labeling in the empirical approach than for the two Ferrenti ones. 
Additionally, substantially fewer principal components are needed for obtaining optimal performance of the ML algorithms for the empirical approach. 
The principal component analyses of the ML methods' performance imply that the Ferrenti and extended Ferrenti approaches, with their strong focus on band gap and bonding character when assessing a material's quantum compatibility, largely replicate the specifications of the data labeling process. 
Valuable insight is gained from the empirical approach which highlights the importance of symmetry related properties in the bond orientations and wavefunctions. 

The findings presented here firmly establish that material informatics (from data mining via featurization to machine learning predictions) is a viable and important route to new discoveries in important fields.  Our focus has been on predicting new candidate materials to host single-photon emitters and spin centers for quantum technology applications, but the framework we have developed is suitable for other fields as well. We suggest two possible paths to further exploit our findings. One aspect is the pursuit of experimental verification of QT compatible effects in the materials predicted as suitable by the four ML methods. Another, perhaps even more important, route is to use the features and trends identified during the data mining and prediction process to understand the distinct material characteristics that enable quantum effects to manifest, opening  thereby up for new discoveries in the field of quantum technologies.

\section*{Methods}

\subsection*{Databases}
The Materials Project \cite{Jain2013, Jain2018} is an open-source project containing ground state properties of materials calculated using density functional theory (DFT) as implemented in the Vienna Ab initio Simulation Package (VASP) \cite{Kresse1996}. The Perdew-Burke-Ernzerhof \cite{Perdew1996} (PBE) functional is used to calculate band structures, while for transition metals, a $+U$ correction is applied to correct for correlation effects \cite{Wang2006}. The project is known as the initiator of materials genomics and offers a variety of calculated properties for over one hundred thousand inorganic crystalline materials, with frequent updates and extensions. Data extraction from Materials Project was performed in December of $2020$ for the Ferrenti and extended Ferrenti approaches, and in March $2021$ for the empirical approach. Therefore, the initial dataset for the two former approaches includes $77$ more materials than that for the empirical approach due to erroneous entries that have been removed from the Materials Project database.

The Open Quantum Materials Database (OQMD) \cite{Saal2013, Kirklin2015} contains thermodynamic and structural properties of more than $600.000$  materials. The calculations are performed with the VASP software and the electron exchange and correlation are described with the PBE functional. The $+U$ extension is included for several calculations considering specific elements \cite{Stevanovic2012}. Data extraction from OQMD was done in February of $2021$. 

JARVIS-DFT \cite{Choudhary2020} is an open-source database based on the VASP software and consists of roughly $40.000$ three-dimensional materials using the vdW-DF-OptB88 (OPT) functional \cite{Thonhauser2007, Klimes2011}. Structures included in the database are originally taken from the Materials Project \cite{Jain2013, Jain2018}, and then re-optimized using the OPT functional. Finally, the combination of the OPT and modified Becke-Johnson (mBJ) functionals \cite{Tran2009} is used to obtain a representative band gap for each structure \cite{Choudhary2018a}. Data extraction from JARVIS-DFT was done in January of $2021$, were we utilized the version made available on 30.04.2021 (see Ref. \cite{Choudhary2020}).

The AFLOW \cite{Curtarolo2012, Curtarolo2012a, Calderon2015} repository is an automatic software framework for the calculations of a wide range of inorganic material properties. They utilize the PBE functional within VASP to relax and optimize the structure from the ICSD. Data extraction was performed in the period of January to February of $2021$.

AFLOW-ML \cite{Isayev2017} is an application programming interface (API) that uses machine learning to predict thermo-mechanical and electronic properties based on the chemical composition and atomic structure alone, which are denoted as \textit{fragment descriptors}. Initially, the API decides whether a given material is a metal or an insulator, where the latter is confirmed with an additional regression method to predict the band gap. The accuracy is validated by a five-fold cross-validation process for each ML method, where they report a $93 \ \%$ prediction success of their initial binary classification method. In this work we utilized the Property Labeled Material Fragments (PLMF) openly available at their website 
(see Ref. \cite{Isayev2017}).
We extract the crystallographic information files (CIF) for the crystals from Materials Project, use the CIF files as input to AFLOW-ML, which then returns an anticipated band gap. This process was executed during January of $2021$. 

\subsection*{Material Informatics}  
Matminer \cite{Ward2018} is an open source toolkit for material analysis written in Python. Matminer provides modules to extract information from a wide variety of databases. Additionally, they provide the tools to construct possibly thousands of features from calculations based on a material's composition, structure and electronic properties from DFT calculations, and have frameworks for visualization and automatic machine learning. 
To apply Matminer's featurization tools, we extend an existing implementation by \citeauthor{Breuck2021} \cite{Breuck2021}, which was used to generate a supervised machine learning framework called the MODnet. The implementation by \citeauthor{Breuck2021} provides featurization for a material's composition, structure and atomic sites. However, Matminer also provides featurization tools for a material's density of states (DOS) and band structure. Therefore, we extend their implementation to facilitate such featurizations. The features selected for featurization herein are summarized in the Supplementary Information at \cite{supplementary}. 

Pymatgen, a robust and open-source Python library for material analysis \cite{pymatgen}, was also employed to extract and generate features for several of the databases mentioned above. 

\subsection*{Machine Learning} 

Machine learning represents the science of giving computers the ability to learn without being explicitly programmed. The idea is that generic algorithms exist which can be used to find patterns in a broad class of datasets without having to write code specifically for each problem. The algorithm builds its own logic based on the data. 

The approaches to machine learning are many, but are often split into two main categories: supervised and unsupervised. In supervised learning we know the answer to a problem, and let the computer deduce the logic behind it. On the other hand, unsupervised learning is a method for identifying patterns and relationships in datasets without any prior knowledge of the system. Many researchers also operate with a third category, namely reinforcement learning. This is a paradigm of learning inspired by behavioral psychology, where learning is achieved by trial-and-error, solely from rewards and punishment. In this work our focus is on supervised learning only with labeled data for classification problems.

In this work we have applied four well-known and tested ML methods for classification problems, these are (see for example \cite{Hastie2009,Mehta2019} for discussions and applications):
\begin{enumerate}
    \item Logistic regression,
    \item Decision trees,
    \item Random forests,
    \item Gradient boosting.
\end{enumerate}
Logistic regression \cite{Hastie2009} is a simple and frequently used method for binary and multi-category classification problems. In addition to logistic regression, we have also applied and tested the predictions made by decision trees and ensemble methods like random forests and gradient boosting, the latter through the application of the computationally efficient XGBoost library \cite{xgboost2016}. Gradient boosting and random forests use decision trees as weak learners and improve their predictability. For random forests this is implemented through a collection of randomized decision trees where a subset of the features in the datasets are selected randomly when building a decision tree. Boosting methods like gradient boosting use decision trees as weak learners and improve upon these by an iterative process that involves the estimation of the gradients of the cost/loss function  \cite{Hastie2009}. Pure decision trees can easily lead to overfitting of the data under study, leading to a ML method that exhibits a high variance. Ensemble methods like random forests and gradient boosting on the other hand tend to soften the overfitting problem, resulting in both a small bias and a reduced variance of the employed method, see for example Refs.~\cite{Hastie2009,Mehta2019} for an in-depth discussion of the bias-variance trade-off in machine learning. Gradient boosting implemented through the  XGBoost library \cite{xgboost2016} is widely used by data scientists to achieve state-of-the-art results on many machine learning challenges. 

\section*{Data availability} 
The datasets that support the findings of this study are available online at \cite{Ohebbi2021}.

\section*{Code availability} 
The codes employed to develop the machine learning results are available online at \cite{Ohebbi2021}. 

\bibliography{apssamp}

\begin{acknowledgments}

The work of LV and MEB was supported by the Research Council of Norway and the University of Oslo through the frontier research projects FUNDAMeNT (no. 251131) and QuTe (no. 325573). 
The work of MEB was supported by an ETH Zurich Postdoctoral Fellowship. 
The work of MHJ was supported by the U.S. Department of Energy, 
Office of Science, office of Nuclear Physics under grant 
No. DE-SC0021152 and U.S. National Science Foundation Grants
No. PHY-1404159 and PHY-2013047. 
The work of SGWL and ØSS was supported by the Norwegian Directorate for International Cooperation and Quality Enhancement in Higher Education (DIKU) which supports the Center for Computing in Science Education (CCSE).

\end{acknowledgments}

\section*{Author contributions}
MEB, LV and MHJ conceived the main theme of the project. OH developed the programs and performed the bulk of the work while MEB lead the writing process. All authors have contributed to the writing of the paper and the discussion and analysis of the data.

\section*{Competing interests}
The authors declare no competing interests.

\section*{Additional information}
Correspondence and requests for materials should be addressed to \url{bathen@aps.ee.ethz.ch}. Supplementary material is available at \url{https://github.com/mhjensen/PredictingSolidStateQubitCandidates}.

\end{document}